\documentclass[prl,aps,amssymb,twocolumn,superscriptaddress,notitlepage]{revtex4-1}
\usepackage{amsmath}
\usepackage{amssymb}
\usepackage{amsthm}
\usepackage{amsfonts}
\usepackage{listings}
\usepackage{enumerate}
\usepackage{latexsym}
\usepackage{graphicx}
\usepackage[caption=false]{subfig}
\usepackage{blkarray}
\usepackage{array}
\usepackage{color}
\usepackage[normalem]{ulem}
\usepackage{hyperref}

\makeatletter
\DeclareRobustCommand{\element}[1]{\@element#1\@nil}
\def\@element#1#2\@nil{%
  #1%
  \if\relax#2\relax\else\MakeLowercase{#2}\fi}
\makeatother

\newcommand{\SUPPpxyHam}{Sec. S1} %{sec:pxyHam}
\newcommand{\SUPPtciphase}{Sec. S1A} %{sec:tciphase}
\newcommand{\SUPPpxyirreps}{Sec. S1B} %{sec:pxyirreps} 
\newcommand{\SUPPSOCxy}{Sec. S1C} %{sec:SOCxy}
\newcommand{\SUPPsymmetries}{Sec. S2} %{sec:208symmetries} 
\newcommand{\SUPPwilsongapped}{Sec. S3A} %{sec:208wilsongapped} 
\newcommand{\SUPPwilsoncrossing}{Sec. S3B} %{sec:208wilsoncrossing}
\newcommand{\SUPPwilsofwils}{Sec. S4} %{sec:wilsofwils}

\begin{document}
\title{Topology of Disconnected Elementary Band Representations}
\author{Jennifer Cano}
\affiliation{Princeton Center for Theoretical Science, Princeton University, Princeton, New Jersey 08544, USA}
\author{Barry Bradlyn}
\affiliation{Princeton Center for Theoretical Science, Princeton University, Princeton, New Jersey 08544, USA}
\author{Zhijun Wang}
\affiliation{Department of Physics, Princeton University, Princeton, New Jersey 08544, USA}
\author{L. Elcoro}
\affiliation{Department of Condensed Matter Physics, University of the Basque Country UPV/EHU, Apartado 644, 48080 Bilbao, Spain}
\author{M.~G. Vergniory}
\affiliation{Donostia International Physics Center, P. Manuel de Lardizabal 4, 20018 Donostia-San Sebasti\'{a}n, Spain}
\affiliation{Department of Applied Physics II, University of the Basque Country UPV/EHU, Apartado 644, 48080 Bilbao, Spain}
\affiliation{Ikerbasque, Basque Foundation for Science, 48013 Bilbao, Spain}
%\affiliation{Max Planck Institute for Solid State Research, Heisenbergstr. 1,
%70569 Stuttgart, Germany.}
\author{C. Felser}
\affiliation{Max Planck Institute for Chemical Physics of Solids, 01187 Dresden, Germany}
\author{M.~I.~Aroyo}
\affiliation{Department of Condensed Matter Physics, University of the Basque Country UPV/EHU, Apartado 644, 48080 Bilbao, Spain}
\author{B. Andrei Bernevig}
\thanks{Permanent Address: Department of Physics, Princeton University, Princeton, New Jersey 08544, USA }
\affiliation{Department of Physics, Princeton University, Princeton, New Jersey 08544, USA}
\affiliation{Donostia International Physics Center, P. Manuel de Lardizabal 4, 20018 Donostia-San Sebasti\'{a}n, Spain}
\affiliation{Laboratoire Pierre Aigrain, Ecole Normale Sup\'{e}rieure-PSL Research University, CNRS, Universit\'{e} Pierre et Marie Curie-Sorbonne Universit\'{e}s, Universit\'{e} Paris Diderot-Sorbonne Paris Cit\'{e}, 24 rue Lhomond, 75231 Paris Cedex 05, France}
\affiliation{Sorbonne Universit\'{e}s, UPMC Univ Paris 06, UMR 7589, LPTHE, F-75005, Paris, France}
\affiliation{LPTMS, CNRS (UMR 8626), Universit\'e Paris-Saclay, 15 rue Georges Cl\'emenceau,\\ 91405 Orsay, France}
\date{\today}
\begin{abstract}

Elementary band representations are the fundamental building blocks of atomic limit band structures.
They have the defining property %\cite{NaturePaper,EBRTheory,GroupTheoryPaper,GraphTheoryPaper,GraphDataPaper,PoFragile}
that at partial filling they cannot be both gapped and trivial.
Here, we give two examples -- one each in a symmorphic and a non-symmorphic space group -- of elementary band representations realized with an energy gap.
In doing so, we explicitly construct a counterexample to a claim by Michel and Zak that single-valued elementary band representations in non-symmorphic space groups with time-reversal symmetry are connected. 
For each example, we construct a topological invariant to explicitly demonstrate that the valence bands are non-trivial.
We discover a new topological invariant: a movable but unremovable Dirac cone in the ``Wilson Hamiltonian'' and a bent-$\mathbb{Z}_2$ index.
\end{abstract}
\maketitle

%\section{Introduction}

The theory of topological quantum chemistry introduced in Ref.~\onlinecite{NaturePaper} diagnoses topological phases based on elementary band representations.
A set of bands is topological if it lacks an ``atomic limit'' that obeys the crystal symmetry (and time-reversal, if desired): formally, an atomic limit exhibits a set of localized, symmetric Wannier functions.\cite{NaturePaper,EBRTheory,PoFragile,Soluyanov2011,Soluyanov2012,Marzari2012,Read2016}
This definition includes all known topological insulating phases.\cite{Kane04,Moore2007,Fu2007,Fu07,Roy2009,BHZ,Teo08,Fu2011,Shiozaki17,Wieder17,Song17}
% including both time-reversal symmetry protected $\mathbb{Z}_2$ topological insulators\cite{Kane04,Moore2007,Fu2007,Fu07,Roy2009,BHZ} and topological crystalline insulators.\cite{Teo08,Fu2011,Shiozaki17,Wieder17,Song17} 
We showed in Refs.~\onlinecite{NaturePaper,EBRTheory} that each atomic limit defines a ``band representation,'' which is a representation of the full space group.
The irreducible representations (irreps) of the little group at each point in the Brillouin zone are completely determined for each band representation.\cite{Zak1980,Zak1982,Bacry1988} 
However, the little group irreps do not define the band representation: two groups of bands can exhibit the same little group irreps but differ by a Berry phase.\cite{Bacry1988b,Bacry1993,Michel1992,EBRTheory}

If a set of bands, separated by an energy gap from all other bands, does not transform as a band representation, it does not have localized, symmetric Wannier functions; consequently, it is topological.\cite{NaturePaper,EBRTheory}
An ``elementary'' band representation (EBR) is not equivalent to a sum of two band representations.
It follows that a disconnected (gapped) elementary band representation must realize a set of topological bands.\cite{NaturePaper,EBRTheory,PoFragile}
Such disconnected EBRs will be the focus of this letter.
All EBRs and their irreps at high-symmetry points in the Brillouin zone can be found on the Bilbao Crystallographic Server.\cite{NaturePaper,GraphTheoryPaper,GraphDataPaper,GroupTheoryPaper,Bilbao1,Bilbao2,Bilbao3}

The theory of topological quantum chemistry also brings to light the different types of trivial-to-topological phase transitions, distinguished by how many symmetry-distinct orbitals contribute to the topological bands.
%or, equivalently, the number of EBRs necessary to describe the full band representation (valence and conduction bands) in the topological phase.
For example, the Kane-Mele model of graphene\cite{Kane04} requires only one type of symmetry-distinct orbital (the two spinful $p_z$ orbitals per unit cell are related by the honeycomb lattice symmetry), 
while the trivial-to-topological transition in HgTe\cite{BHZ} requires both $s$ and $p$ orbitals %, which are not related by symmetry, 
to create a ``band inversion.''
These two types of topological insulators differ in their atomic limit as the distance between atoms is taken to infinity: in the atomic limit of graphene, the band structure consists of a single flat and four-fold degenerate band, corresponding to a single EBR.
In contrast, in HgTe, the atomic limit will consist of two flat bands, one each for the $s$ and $p$ orbitals, corresponding to two distinct EBRs.

In this letter, we will focus on the graphene-like case: topological insulators that derive from a single orbital and its symmetry-related partners.
In the language of band representations, the conduction and valence bands together transform as a single EBR; consequently, either the conduction or valence bands (or both) lack an atomic limit and are topological.\cite{NaturePaper,EBRTheory,PoFragile}
%These types of topological phases do not require a band inversion and can exist with or without spin-orbit coupling (SOC).

We introduce models in a symmorphic and a non-symmorphic space group. %, both with and without SOC.
The symmorphic example describes $p_{x,y}$ orbitals on the honeycomb lattice.
%This describes materials where an energy gap between $p_{x,y}$ and $s$ orbitals prevents the formation of hybridized $sp^2$ orbitals.
Without spin-orbit coupling (SOC), the band structure can be a (gapped) topological crystalline insulator (TCI).
With infinitesimal SOC and time-reversal symmetry, the system exhibits a nontrivial $\mathbb{Z}_2$ index.%, which has been realized experimentally.\cite{Reis2016,Zhou2014}

We were motivated to explore the non-symmorphic example because, as part of their ground-breaking work on the connectivity of energy bands, Michel and Zak conjectured that spinless EBRs in non-symmorphic space groups cannot realize a gapped band structure.\cite{Michel1999,Michel2001} 
%This claim forbids topological insulators in non-symmorphic space groups to arise from a single set of symmetry-related orbitals.
In Ref~\onlinecite{GraphDataPaper}, we explained where Michel and Zak's proof fails.
Here, we pick a particular non-symmorphic space group, $P4_2 32$, and construct a tight-binding model to explicitly show its gapped, topological nature.
In doing so, we find a novel feature: the two-dimensional ``Wilson Hamiltonian'' exhibits a topologically protected band crossing.
%We prove that the Berry phase acquired from winding around the crossing point serves as a topological invariant, which indicates the winding of the Wilson loop on a bent path (a $\mathbb{Z}_2$ generalization of Ref.~\onlinecite{Holler2017}.)

In each example, we derive a bulk topological invariant. % that characterizes the symmetry-protected topological phase.
An essential tool is the ``$k_\parallel$-directed'' Wilson loop, which describes the parallel transport of an isolated set of bands:%, separated from all other bands by an energy gap.
\cite{Zakphase,Fu06,Ryu10,Soluyanov2011,Yu11,Taherinejad14,Alexandradinata14,Hourglass,ArisCohomology,hughesbernevig2016,Wieder17,Holler2017} 
\begin{equation}
 \mathcal{W}_{(k_\perp,k_0)} \equiv  P e^{i\int_{k_0}^{k_0 +2\pi} dk_\parallel A_\parallel(k_\perp,k_\parallel)},  
\label{eq:wilsoncont}
\end{equation}
where $P$ indicates that the integral is path-ordered and $A_\parallel (\mathbf{k})_{ij} = i \langle u_i(\mathbf{k}) | \partial_{k_\parallel} u_j (\mathbf{k})\rangle$ is a matrix whose rows and columns correspond to each eigenstate in the isolated set of bands. 
%$\mathcal{W}_{(k_\perp,k_0)}$ is a function of $k_\perp$.
The eigenvalues of $\mathcal{W}$ are gauge invariant and of the form $e^{i\theta(k_\perp)}$, independent of the ``base point,'' $k_{0}$.\cite{ArisCohomology}
A quantized invariant derived from the Wilson loop is invariant under any deformation of the Hamiltonian that preserves the gap in the spectrum.%, thus providing a topological invariant. 
%When then gap closes, the matrix $A$ ceases to be defined.

%%%%%%%%%%%%%%%%%%%%%%%%%%%%%%
%%%%%%%%%%%%%%%%%%%%%%%%%%%%%%
%%%%%%%%%%%%%%%%%%%%%%%%%%%%%%
%%%%%%%%%%%%%%%%%%%%%%%%%%%%%%
\paragraph{Spinless TCI on the honeycomb lattice}
%\section{Spinless topological phase on the honeycomb lattice}
%\label{sec:spinlessTCI}

\begin{figure}[h]
\centering
\subfloat[]{
	\includegraphics[width=.8in]{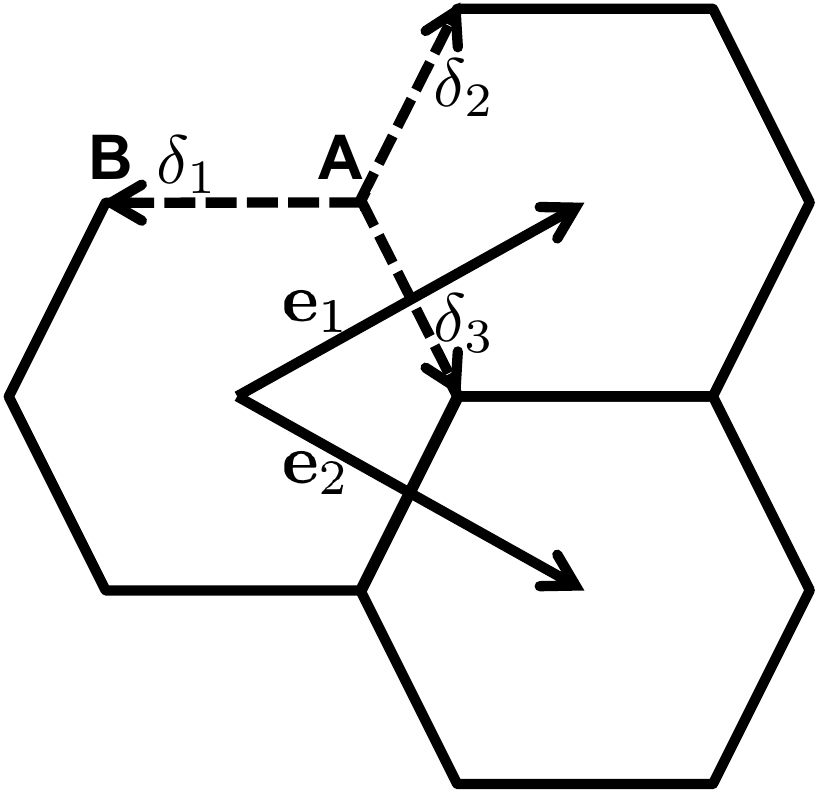}
	\label{fig:graphenebasisvectors}
\quad
	\includegraphics[width=.6in]{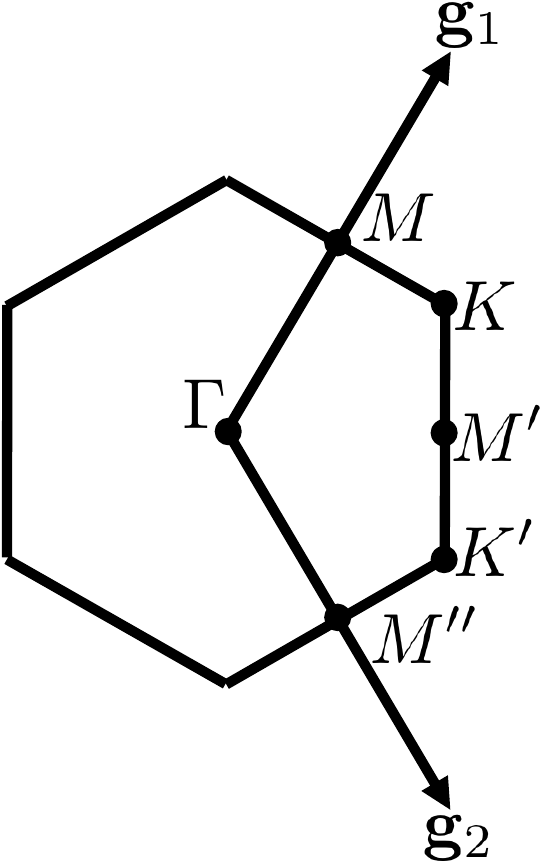}
	%\label{fig:graphenereciprocal}
}
\quad
\subfloat[]{
	\includegraphics[width=1.5in]{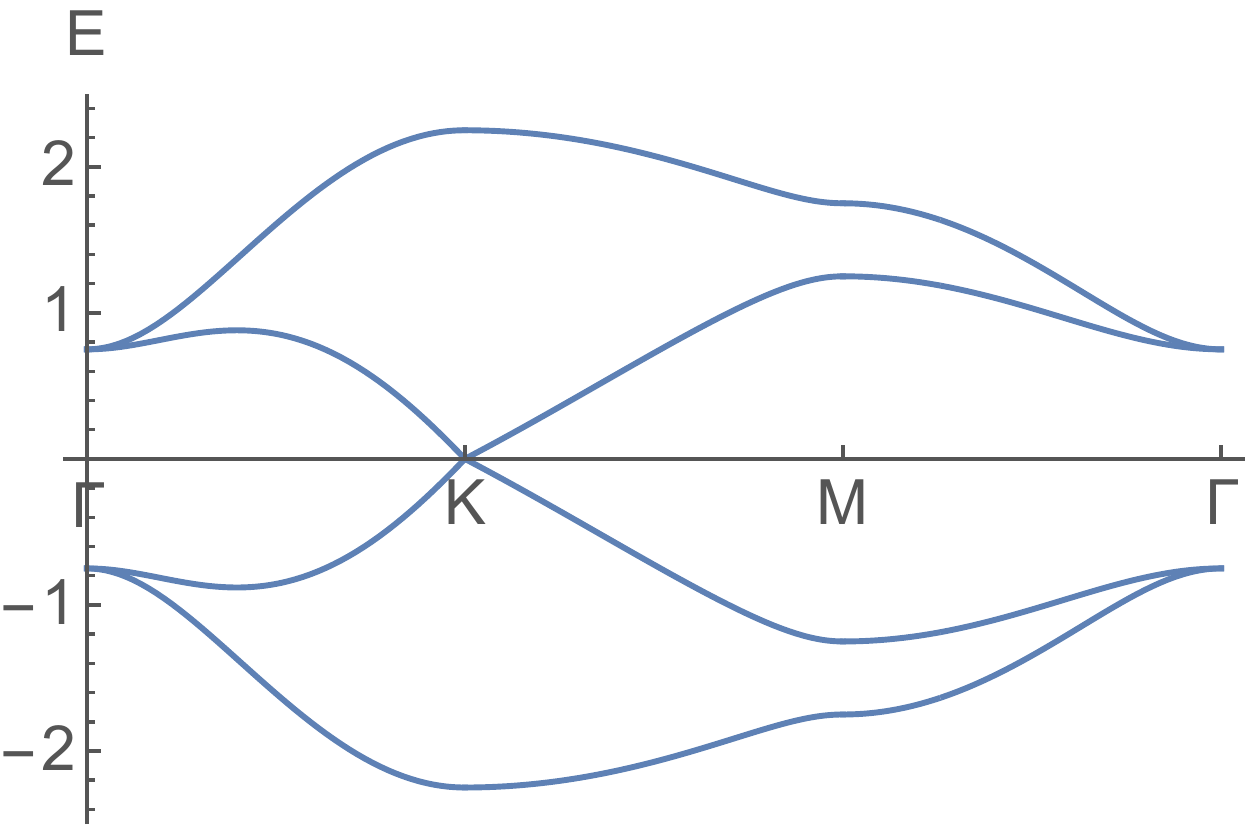}
	\label{fig:pxyspectrum}
}
\quad
\subfloat[]{
	\includegraphics[width=1.5in]{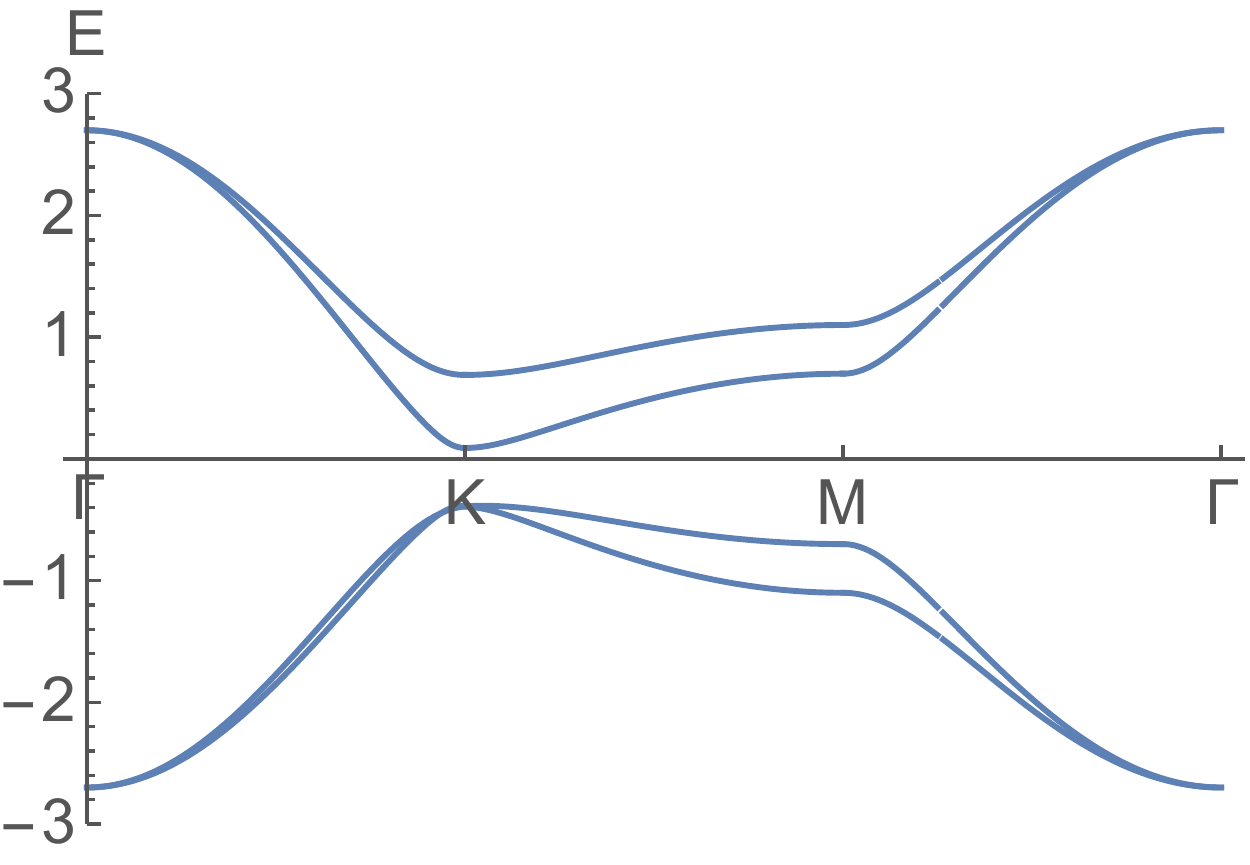}
	\label{fig:pxyspectrum-disc}
} \quad
\subfloat[]{
	\includegraphics[width=1.5in]{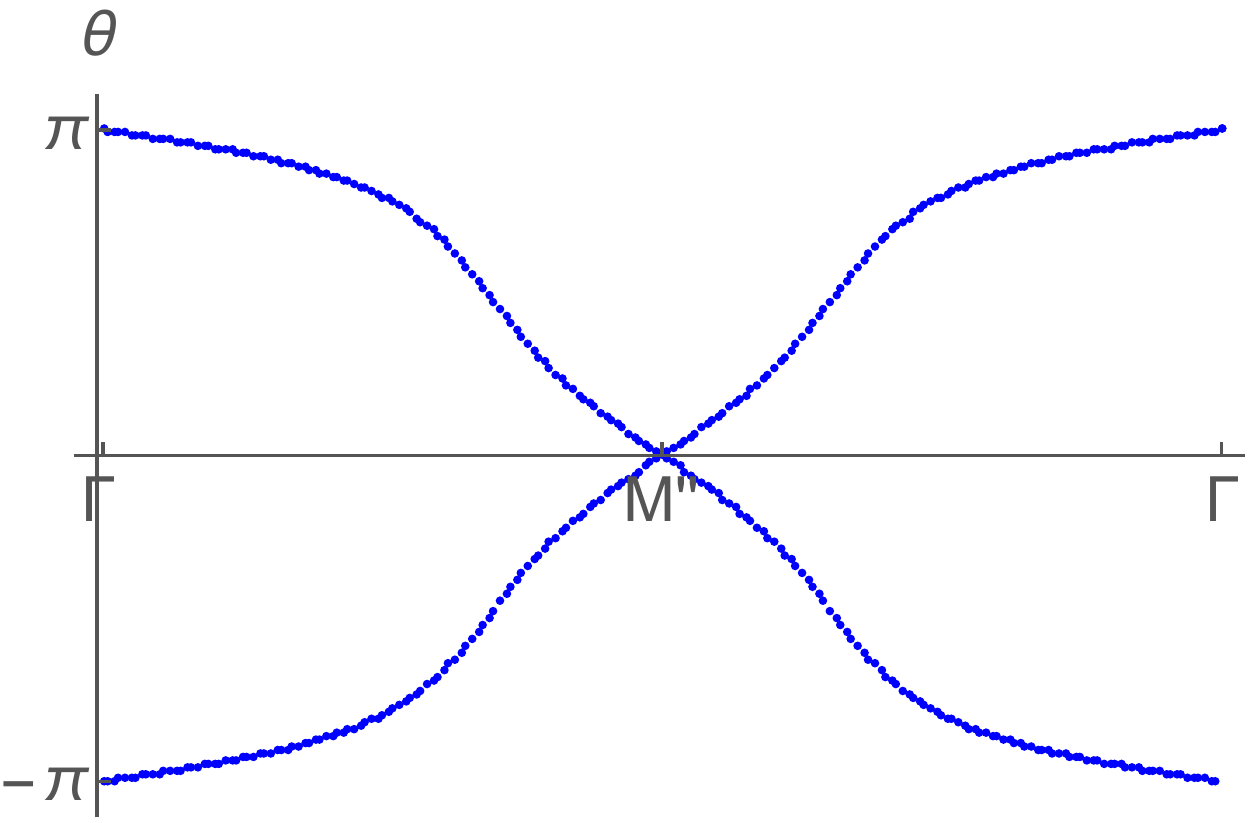}
	\label{fig:pxyWilson}
}
\caption{ (a) Lattice ($\mathbf{e}_{1,2}$) and reciprocal lattice ($\mathbf{g}_{1,2}$) basis vectors. The dotted arrows ($\delta_{1,2,3}$) indicate the vectors between nearest neighbor sites. \textbf{A} and \textbf{B} indicate the sublattices. (b) Spectrum of $H^0_\mathbf{k}$ with $t_\sigma = 1, t_\pi = -.5$
(c) Gapped band structure of $H^0_\mathbf{k}+xH^1_\mathbf{k}$ with $t_\sigma = .8, t_\pi = 1.0, x=.6$ and (d) the argument of its Wilson loop eigenvalues.}
\label{fig:graphenefig}
\end{figure}

%The canonical example of a $\mathbb{Z}_2$ topological insulator is {\blue the $p_z$ orbitals of } graphene with artificially strong next-nearest-neighbor SOC.\cite{Kane04}
We start with spinless $p_{x,y}$ orbitals on the honeycomb lattice,
described by the nearest-neighbor Hamiltonian:
%derived from $\sigma$ and $\pi$ bonds:
\footnote{Eqs.~(\ref{eq:pxyHam}) and (\ref{eq:pxyHamh}) are derived in \SUPPpxyHam.}
%The gapless band structure comprises a connected EBR, which is a symmetry-protected semi-metal at half-filling.
%We then add a symmetry-allowed next-nearest neighbor hopping term that drives the system into a spinless TCI phase.
%In the presence of infinitesimal spin-preserving SOC, this phase becomes a spinful topological insulator. 
%When time-reversal symmetry is enforced, the phase diagram consists of two phases, both of which contain bands with a nontrivial $\mathbb{Z}_2$ index.
%When the gap at half-filling is topological, the model describes the topological phase of bismuthene reported in Ref~\onlinecite{Reis2016}.
%The nearest neighbor Hamiltonian for spinless $p_{x,y}$ orbitals on the honeycomb lattice is given by:
\begin{equation}
H_\mathbf{k}^0  = \begin{pmatrix} 0 & h_\mathbf{k} \\ h_\mathbf{k}^\dagger & 0 \end{pmatrix} 
\label{eq:pxyHam}
\end{equation}
where non-zero blocks mix the $A$ and $B$ sublattices and
\begin{align}
h_\mathbf{k} &= \frac{1}{2} \left( e^{-i\mathbf{k} \cdot \mathbf{\delta}_1} + e^{-i\mathbf{k} \cdot \mathbf{\delta}_2} + e^{-i\mathbf{k} \cdot \mathbf{\delta}_3} \right)(t_\sigma + t_\pi) \mathbb{I} \nonumber\\
&+ \frac{1}{2} \left( e^{-i\mathbf{k} \cdot \mathbf{\delta}_1} -\frac{1}{2} e^{-i\mathbf{k} \cdot \mathbf{\delta}_2} -\frac{1}{2} e^{-i\mathbf{k} \cdot \mathbf{\delta}_3} \right)(t_\sigma - t_\pi)\sigma_z \nonumber\\
&+ \frac{\sqrt{3}}{4} \left(  e^{-i\mathbf{k} \cdot \mathbf{\delta}_2} - e^{-i\mathbf{k} \cdot \mathbf{\delta}_3} \right)(t_\sigma - t_\pi ) \sigma_x
\label{eq:pxyHamh}
\end{align}
The Pauli matrices, $\sigma_{x,y,z}$, act in the $p_{x,y}$ subspace; $t_{\sigma,\pi}$ parameterize $\sigma$ and $\pi$ bond strengths; and $\delta_{1,2,3}$ are the nearest-neighbor vectors (see Fig~\ref{fig:graphenebasisvectors}).
Previously this model with $t_\pi = 0$ was studied for its flat bands.\cite{Wu2007,Liu2013}
The spectrum of $H^0_\mathbf{k}$ is shown in Fig~\ref{fig:pxyspectrum}.
%; it is particle-hole symmetric because $H_\mathbf{k}^0$ anticommutes with $\tau_z \otimes \sigma_0$, where $\tau_z$ acts in the sublattice space.
The degeneracies at $K\equiv \frac{2}{3}\mathbf{g}_1 + \frac{1}{3}\mathbf{g}_2$ and $\Gamma$ are symmetry-required.\cite{Kogan12} 
%They can be looked up on the BCS server using the BANDREP tool\cite{GroupTheoryPaper} for space group $P6mm$, which describes (three-dimensional) stacked layers of graphene.
%the representation of the symmetry group of graphene ($P6mm$) induced from the irrep of $C_{3v}$ corresponding to $p_{x,y}$ orbitals on the honeycomb lattice contains in its subduction onto the little co-group of $K$ the two-dimensional irrep, $K_3$, and contains in its subduction onto the little co-group of $\Gamma$ two two-dimensional irreps, $\Gamma_{5,6}$.
%\footnote{This information can be found on the Bilbao Crystallographic Server using the BANDREP tool for space group $P6mm$ (183), which describes (three-dimensional) stacked layers of graphene.}

To open a gap, we add the following next-nearest neighbor hopping term, which preserves the crystal symmetries of the honeycomb lattice:\footnote{The crystal symmetries of the honeycomb lattice are enumerated in~\SUPPpxyHam.} 
\begin{equation}
H^1_\mathbf{k} = \sin({\textstyle \frac{1}{2}}\mathbf{k}\cdot \mathbf{e}_1)\sin({\textstyle \frac{1}{2}}\mathbf{k} \cdot \mathbf{e}_2) \sin( {\textstyle \frac{1}{2}}\mathbf{k} \cdot (\mathbf{e}_1 - \mathbf{e}_2) ) \tau_z \otimes \sigma_y,
\label{eq:pxynext}
\end{equation}
where the matrices $\tau_i$ act in the sublattice subspace.
The term in Eq.~(\ref{eq:pxynext}) changes the energy-ordering of the bands at $K$, while preserving the two-fold degeneracy.
For large enough $|x|$, $H_\mathbf{k}^0 +x H_\mathbf{k}^1$ can be gapped, as in Fig.~\ref{fig:pxyspectrum-disc};
see \SUPPtciphase\ for a phase diagram.%we compute the phase diagram as a function of $x$ and $t_{\sigma,\pi}$.

%\begin{figure}[h]
%\centering
%\subfloat[]{
%	\includegraphics[width=1.5in]{pxpyHam-disc.pdf}
%	\label{fig:pxyspectrum-disc}
%} \quad
%\subfloat[]{
%	\includegraphics[width=1.5in]{pxpy-Wilson.pdf}
%	\label{fig:pxyWilson}
%}
%\caption{(a) Gapped band structure of $H^0_\mathbf{k}+xH^1_\mathbf{k}$ with $t_\sigma = .8, t_\pi = 1.0, x=.6$ and (b) the argument of its Wilson loop eigenvalues.}
%\end{figure}

The spectrum in Fig~\ref{fig:pxyspectrum-disc} represents a disconnected EBR.\cite{NaturePaper,EBRTheory} 
We construct a non-trivial bulk topological invariant from the $\mathbf{g}_1$-directed Wilson loop of the lower two bands. % (the same would hold for the upper two bands.)
Its eigenvalues are shown in Fig.~\ref{fig:pxyWilson} as a function of the base point.
When the base point is $\Gamma$ or $M$, the Wilson loop eigenvalues ($-1$ and $+1$, respectively\footnote{\SUPPtciphase\ shows the possible $C_{2z}$ eigenvalues.}) are completely determined by the $C_{2z}$ eigenvalues\cite{Alexandradinata14,Holler2017}
(the $C_{2z}$ operator is $-\tau_x \otimes \sigma_0$.\footnote{The crystal symmetries of the honeycomb lattice are enumerated in~\SUPPpxyHam.} )
This forces the ``Wilson bands'' to wind in opposite directions.
%The eigenvalues of the Wilson loop that starts at $\Gamma$ and passes through $M$ are equal to $-1$ because the two occupied bands at $\Gamma$ have the same $C_{2z}$ eigenvalues, which are opposite to those at $M$ (this is always true in the gapped phase; see \SUPPtciphase.)\cite{Alexandradinata14} The eigenvalues of the Wilson loop that starts at $M''$ and passes through $M'$ are both equal to $+1$ because the occupied bands at $M''$ and $M'$ have the same $C_{2z}$ eigenvalues.
%In Fig~\ref{fig:pxyWilson}, we plot the eigenvalues of the $\mathbf{g}_1$-directed Wilson loop as a function of its base point; the ``Wilson bands'' are forced to wind in opposite directions because of their eigenvalues at $\Gamma$ and $M''$.
The quantized eigenvalues at $\Gamma$ and $M$ prevent the Wilson spectrum from being smoothly deformed to flat, which indicates that the valence bands are topologically nontrivial.

The Wilson loop winding requires that both occupied bands of $H^0_\mathbf{k} + xH^1_\mathbf{k}$ at $\Gamma$ have the same $C_{2z}$ eigenvalue, $\eta$, and that both occupied bands at $M$ have the $C_{2z}$ eigenvalue $-\eta$.
Consider the Wilson loop of three bands: the two occupied bands and a third, trivial, band, not in our model.
If the $C_{2z}$ eigenvalues of the third band at $\Gamma$ and $M$ are both equal to $\eta$,
% the same at $\Gamma$ as at $M$ and equal to the $C_{2z}$ eigenvalues at $\Gamma$ of the lower two bands of $H_\mathbf{k}^0$, 
then the eigenvalues of the three-band Wilson loop will not be quantized at $M$ and it will fail to wind.
Thus, the topological invariant is not stable to adding a third band to the projector (although the winding of the projector onto two bands \textit{is} invariant under adding a third band as long as the gap between the third band and the existing bands does not close.) %, i.e., as long as the two-band projector remains well-defined.)
The existence of a topological invariant that depends on the number of bands is reminiscent of the ``Hopf insulator.''\cite{Moore2008}

\paragraph{Spinful topological phases}

\begin{figure}[h]
\centering
\subfloat[]{
	\includegraphics[width=1.5in]{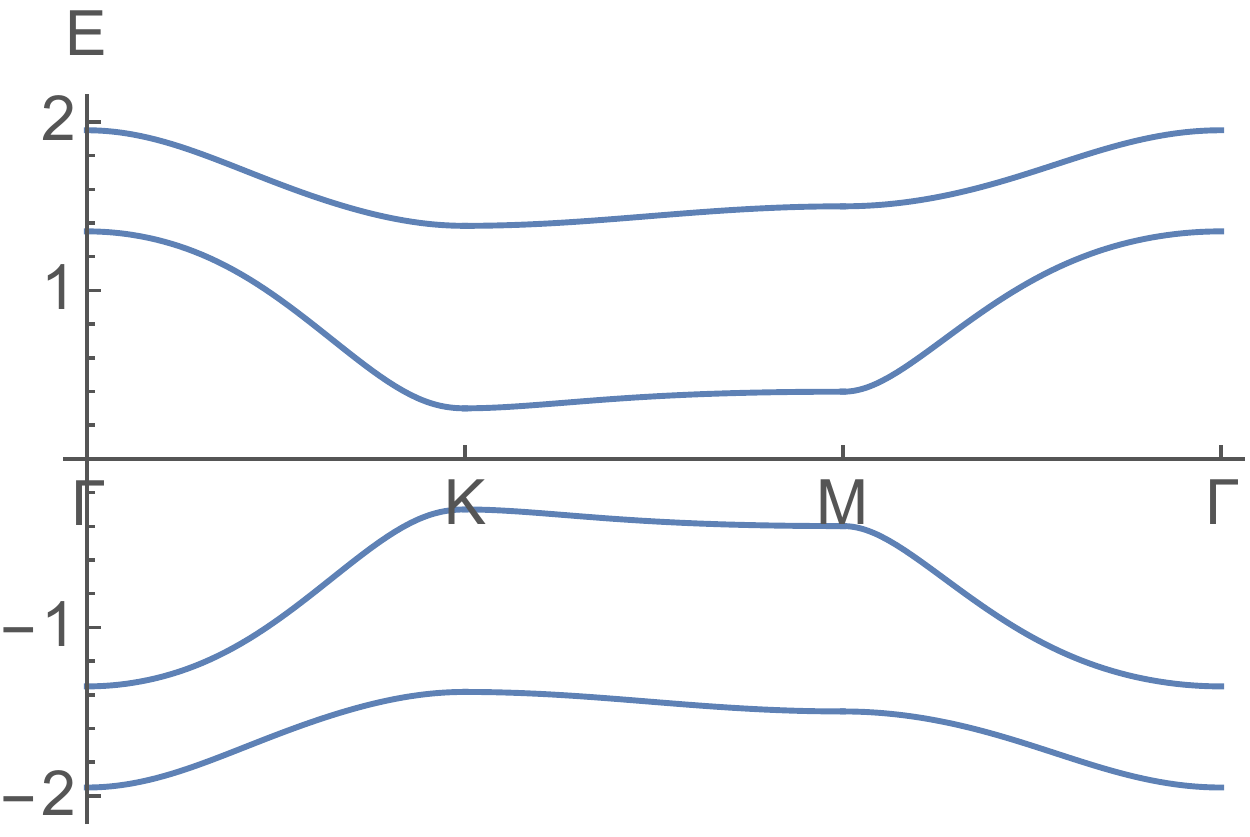}
	\label{fig:pxpy-alltop}
} \quad
\subfloat[]{
	\includegraphics[width=1.5in]{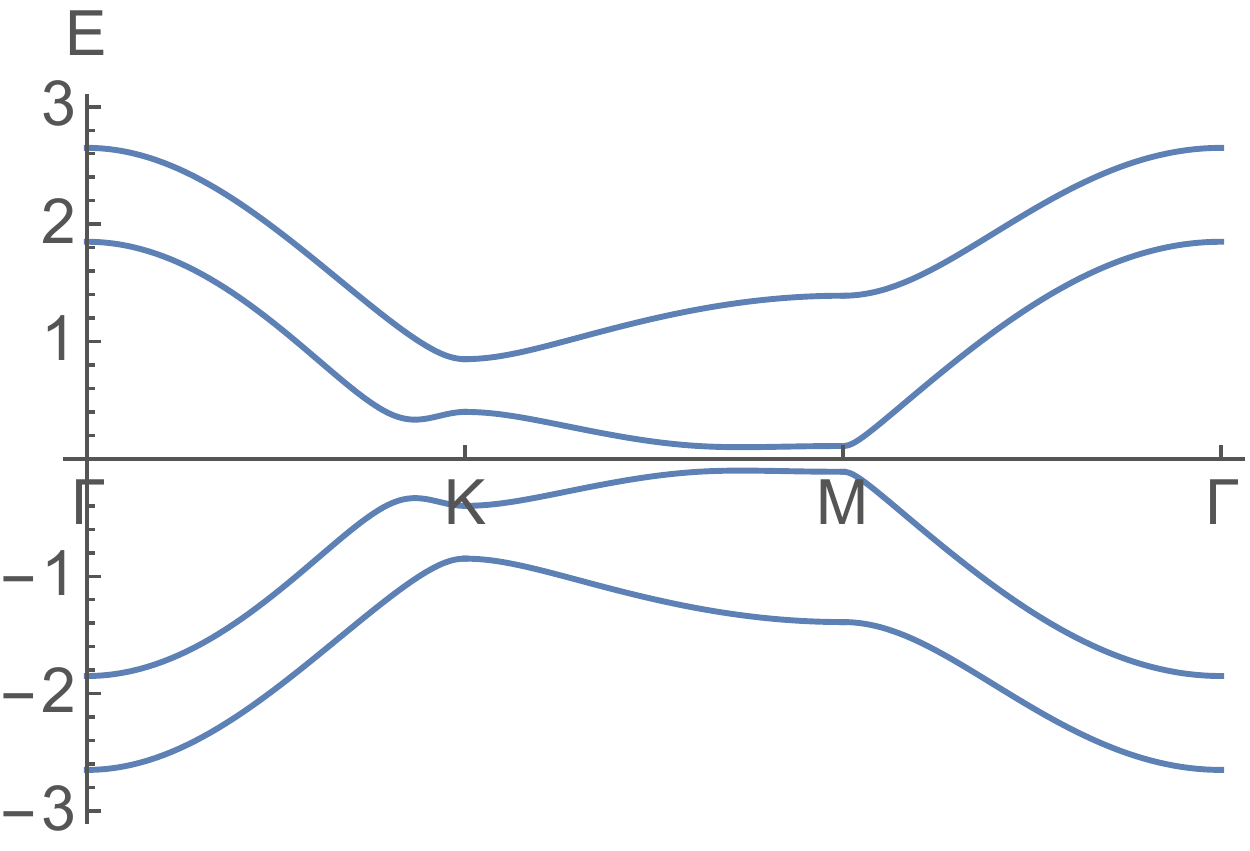}
	\label{fig:pxpy-sometop}
	}
\quad
\subfloat[]{
	\includegraphics[width=1.5in]{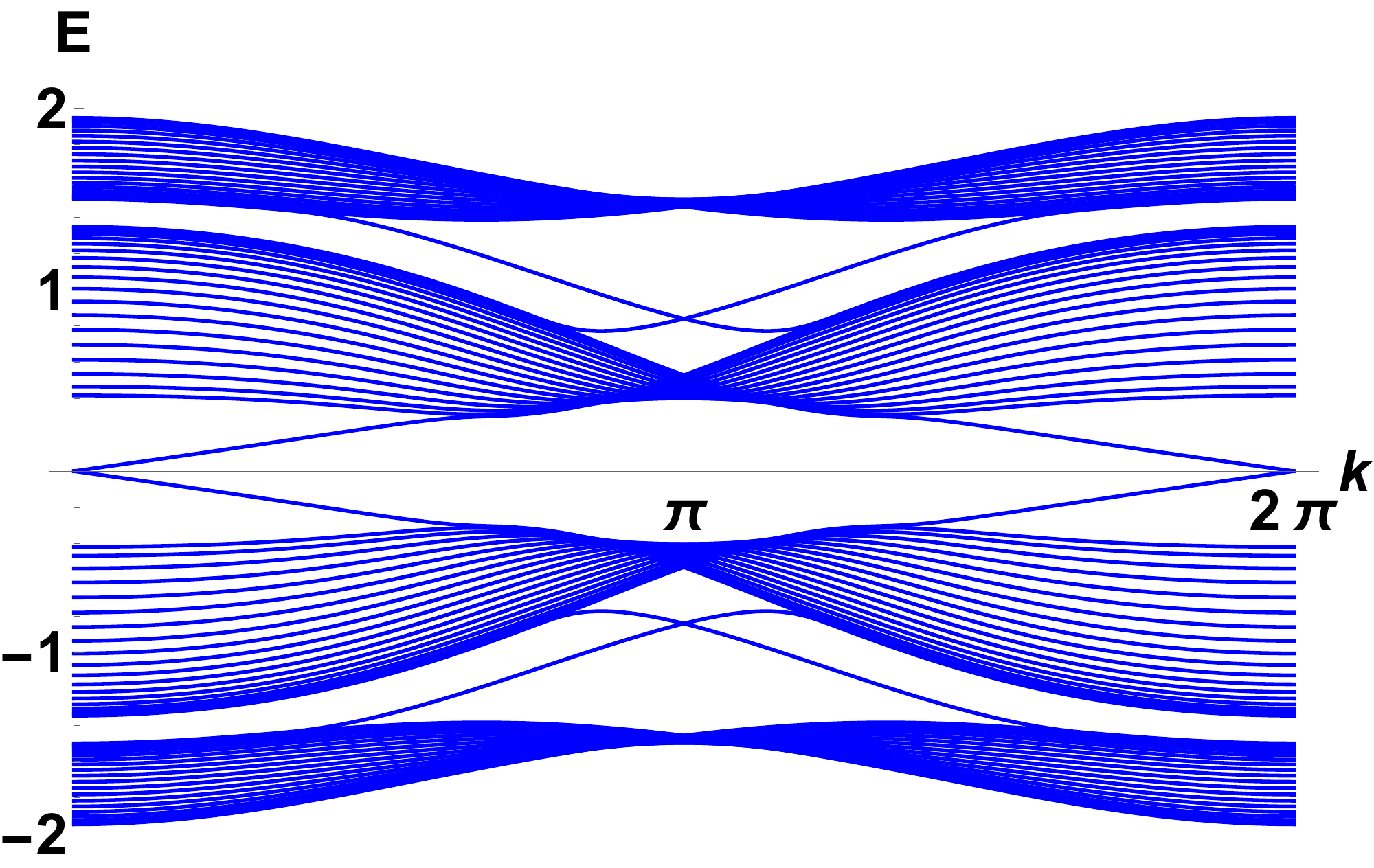}
	\label{fig:pxpy-alltop-slab}
} \quad
\subfloat[]{
	\includegraphics[width=1.5in]{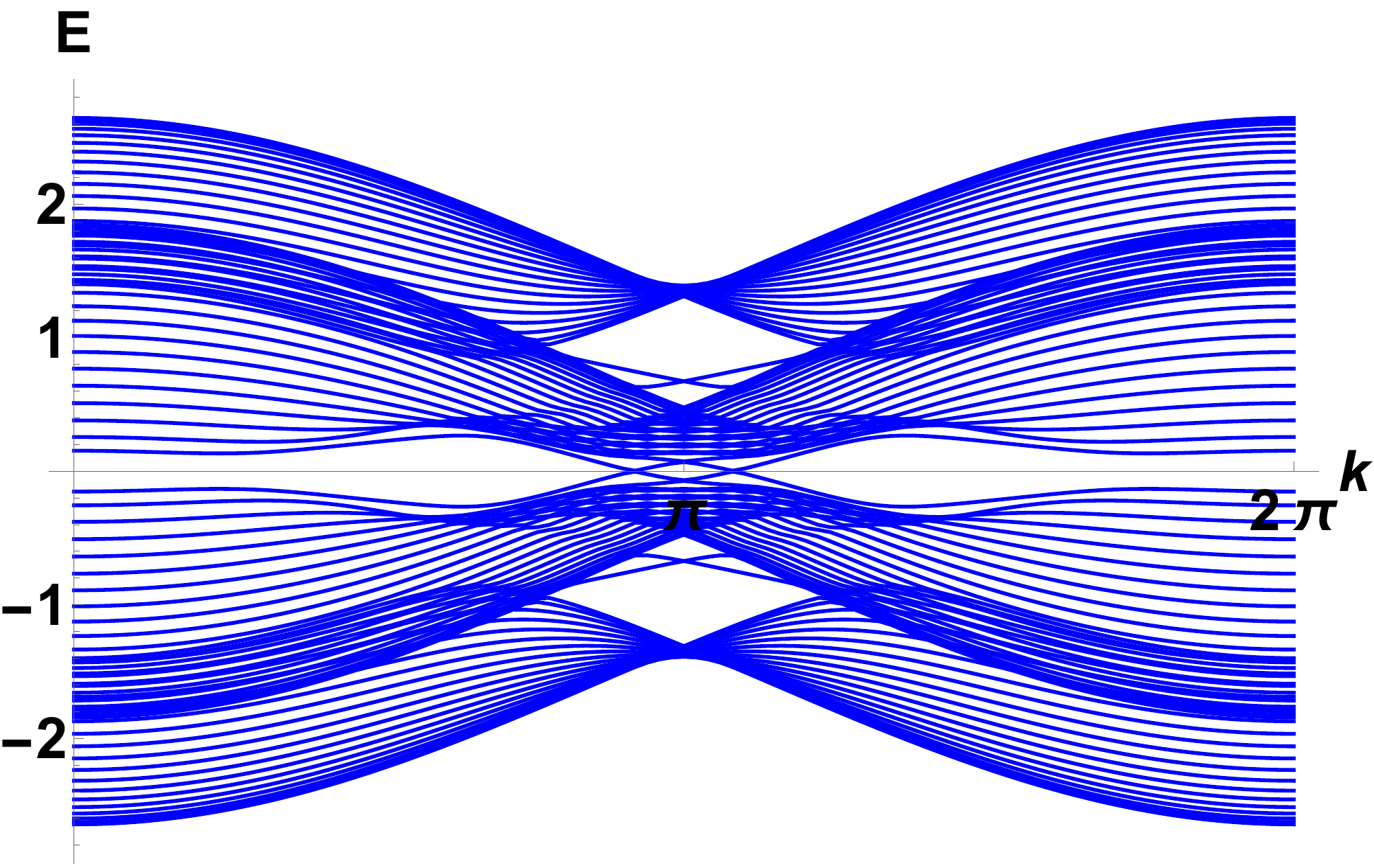}
	\begin{picture}(0,0)
\put(-40,40){\includegraphics[height=1cm]{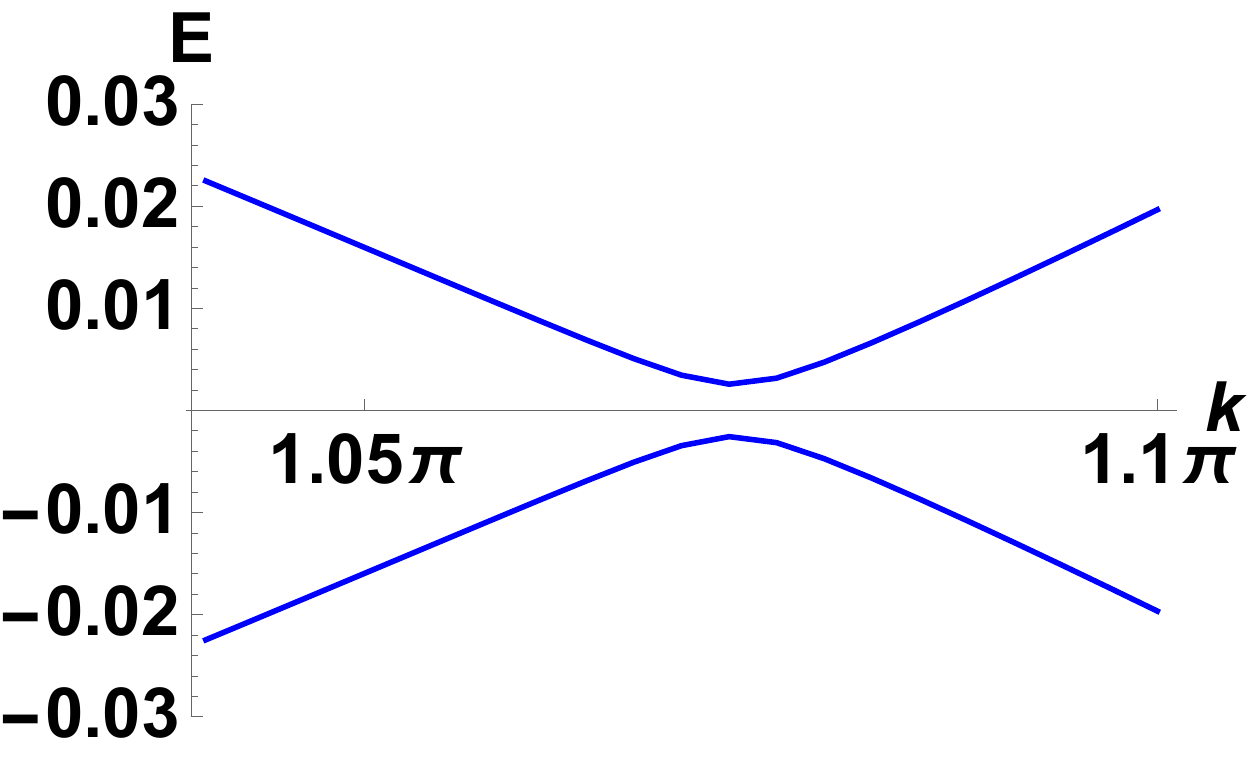}}
\end{picture}
	\label{fig:pxpy-sometop-slab}
}
\caption{Band structures with an onsite $L\cdot S$ term for (a,b) periodic and (c,d) slab boundary conditions. This SOC term preserves inversion symmetry, such that all bands remain doubly degenerate, but is general enough to open all possible gaps in the band structure.
In (a) only the lowest and highest bands have a nontrivial $\mathbb{Z}_2$ index and hence the slab boundary conditions in (c) reveal edge states in all three gaps.
In (b) all four bands have a nontrivial $\mathbb{Z}_2$ index and hence the slab boundary conditions in (d) reveal edge states in only the upper and lower gaps. %, although the edge states merge with the bulk states. 
The inset in (d) resolves the avoided crossing at $E=0$.}
\label{fig:pxySOC}
\end{figure}

We now consider SOC.
Spinful $p_{x,y}$ orbitals decompose into three irreps of the site-symmetry group. % ($C_{3v}$)
%: one 2D irrep (the Kramers pair $|S=\frac{1}{2}, S_z=\pm \frac{1}{2}\rangle$) and two 1D irreps ($|S=\frac{3}{2},S_z=\frac{3}{2}\rangle$ and $|S=\frac{3}{2},S_z=-\frac{3}{2}\rangle$) that are time-reversed partners. 
Bands derived from these three irreps transform as a sum of three EBRs,\cite{NaturePaper,EBRTheory}
which generically split into four sets of disconnected bands, as in Figs~\ref{fig:pxpy-alltop} and \ref{fig:pxpy-sometop}.
At least one set of disconnected bands is either an obstructed atomic limit -- it can be adiabatically deformed to a Hamiltonian comprised of orbitals that reside at the center of the hexagon rather than the corners -- or a topological band that does not have any atomic limit.\footnote{The possibilities are described in \SUPPSOCxy.}

If time-reversal symmetry is enforced, we can consider the $\mathbb{Z}_2$ index.
For small spin-conserving SOC that does not invert the bands at $\Gamma$ or $M$, the $C_{2z}$ eigenvalues in the spinless phase determine the $\mathbb{Z}_2$ index of each set of bands (conserving spin amounts to enforcing inversion symmetry.)
Our simple, but physically motivated, model yields two phases, shown in Fig~\ref{fig:pxySOC}: either all three or the first/third gaps are $\mathbb{Z}_2$ topological, while the middle gap is not; there is no phase in which all gaps are $\mathbb{Z}_2$ trivial.\footnote{The possible $C_{2z}$ eigenvalues are computed in \SUPPSOCxy .}
%This is confirmed by plotting the slab band structure of $H_\mathbf{k}^0$ with an onsite $L\cdot S$ term in Fig~\ref{fig:pxySOC}, which reveals edge states in either one or three of the gaps separating bulk states.
We show in \SUPPSOCxy\ that only spin-conserving SOC can open a gap in the spinless band structure; hence, if non-spin-conserving SOC is present and does not invert any bands, it will alter the band structure but not change the $\mathbb{Z}_2$ index.

\paragraph{Material realization}
%\label{sec:materials}

%Since bands derived from $p_{x,y}$ orbitals on the honeycomb lattice transform as a single EBR, their band structure will either be a symmetry-protected semi-metal or a topological insulator.
%We have provided a Hamiltonian that realizes both of these phases.
The spinless semi-metallic model $H_\mathbf{k}^0$ consists of nearest-neighbor Slater-Koster\cite{SlaterKoster} terms; thus, it is widely applicable to two-dimensional planar honeycomb systems.
To exhibit the TCI phase, the next-nearest neighbor term, $H_\mathbf{k}^1$, must be dominant in order to open a gap.
The relative strength of the hopping terms varies with strain or buckling.
%This could occur in a buckled honeycomb system (where inversion plays the role of $C_{2z}$) because the buckling geometry enhances the relative strength of the next-nearest neighbor coupling.

The non-trivial $\mathbb{Z}_2$ phases will be present whenever SOC is large enough to open an observable gap, but not so large to invert the bands at $\Gamma$ and $M$.
In particular, $H_\mathbf{k}^0$ with SOC describes bismuth grown on an SiC substrate, 
%because the $p_z$ orbitals bond strongly to the substrate, the bands at the Fermi level are predominantly of $p_{x,y}$ character.\cite{Reis2016,Zhou2014}
%This is 
consistent with the topological edge states reported in Ref~\onlinecite{Reis2016}.

%%%%%%%%%%%%%%%%%%%%%%%%%%%%%%
%%%%%%%%%%%%%%%%%%%%%%%%%%%%%%
%%%%%%%%%%%%%%%%%%%%%%%%%%%%%%
%%%%%%%%%%%%%%%%%%%%%%%%%%%%%%
\paragraph{Non-symmorphic gapped EBR}

\begin{figure}[t]
\centering
%\subfloat[]{
%	\includegraphics[width=1.5in]{SG208-nnvecs.png}\label{fig:208unitcell}
%	}\quad
%\subfloat[]{
%	\includegraphics[width=1.5in]{SG208-bands-deg.pdf}\label{fig:208bands-deg}
%	}\quad
\subfloat[]{
	\includegraphics[width=1.4in]{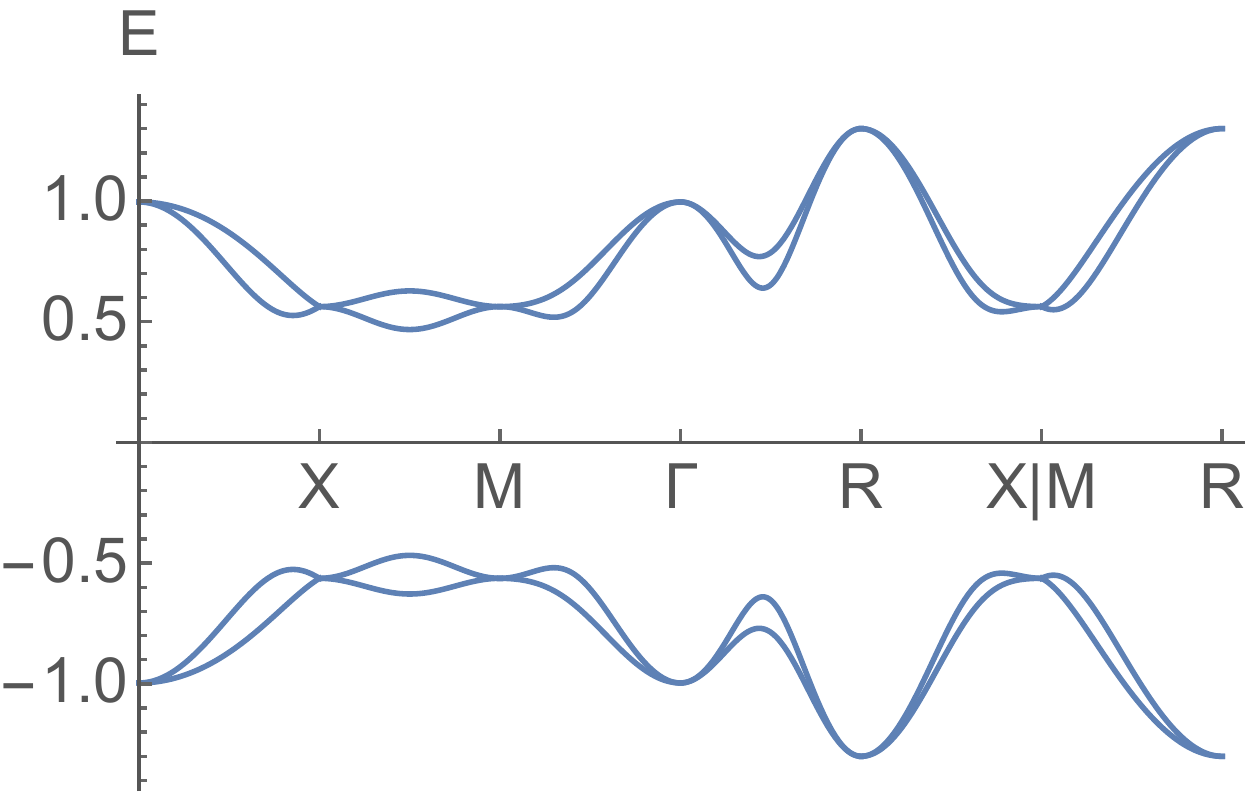}\label{fig:208bands-nodeg}
	\put(-5,30){\includegraphics[height=1.2cm]{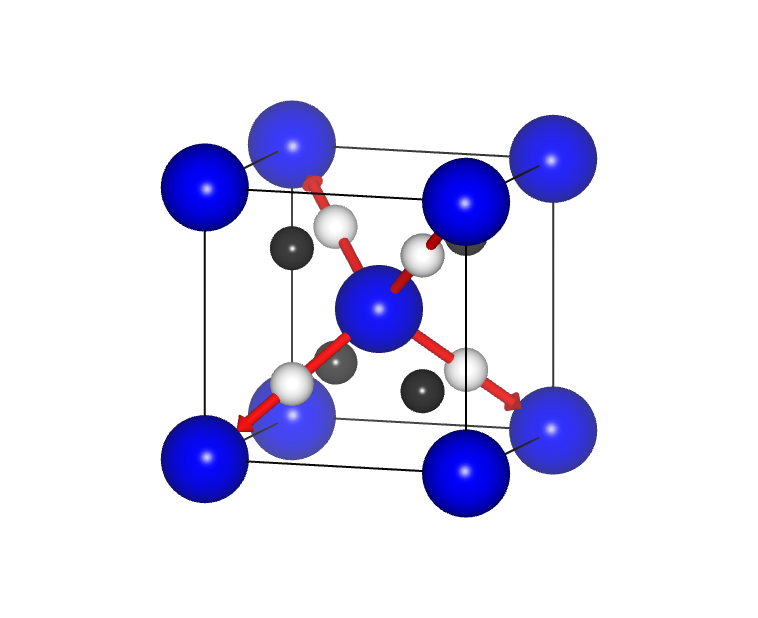}}
	}\quad\quad\quad
\subfloat[]{
	\includegraphics[width=1.4in]{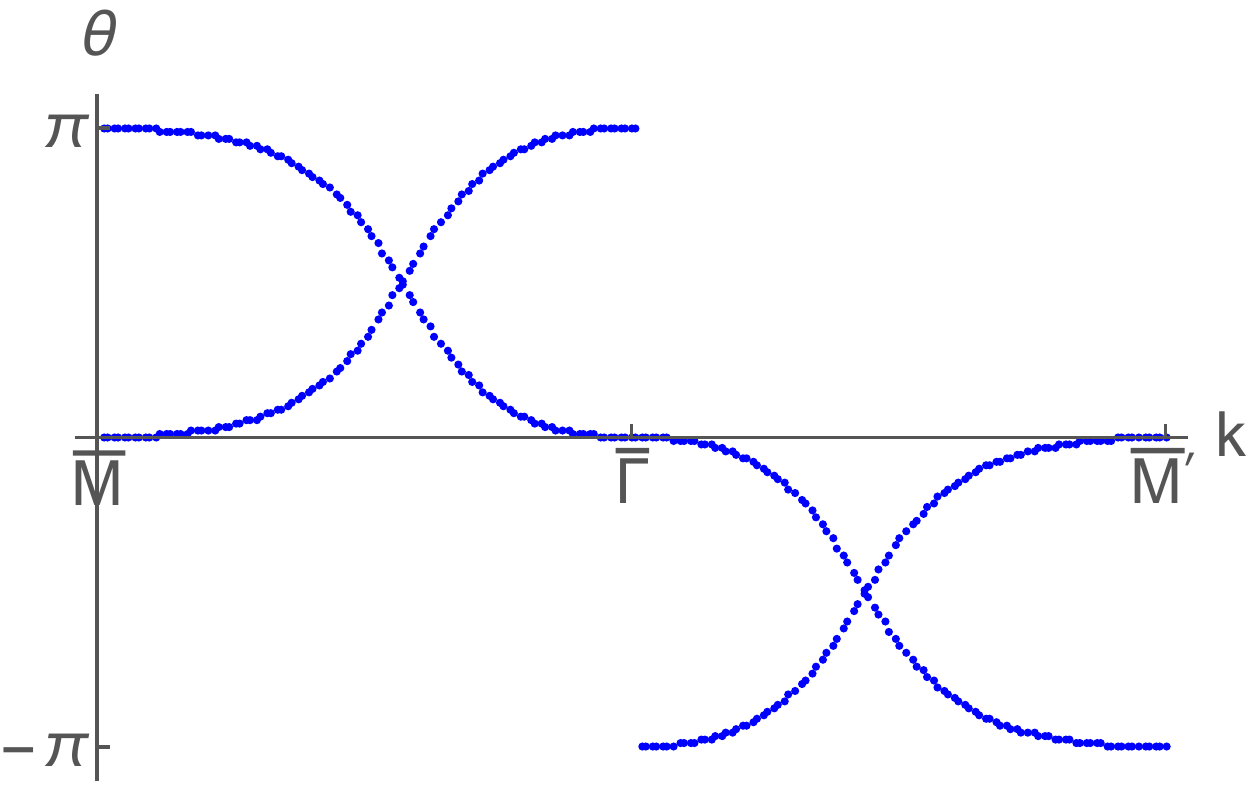}\label{fig:208wilsonz}
	\begin{picture}(0,0)
\put(-35,35){\includegraphics[height=1.2cm]{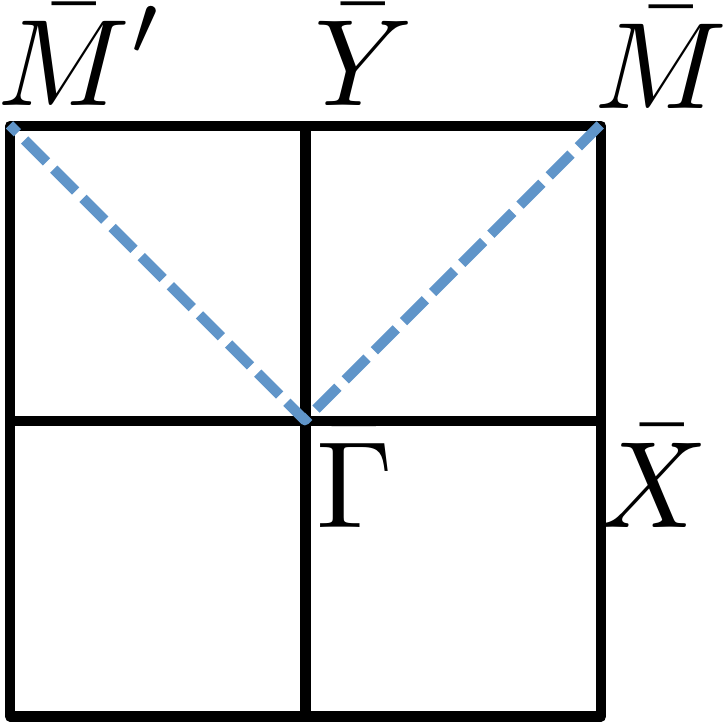}}
\end{picture}
	}
\caption{%(a) Unit cell. Orbitals on the blue atoms enter the Hamiltonian; the white/black atoms create a crystal field with the symmetry of $P4_232$.
%Red arrows indicate the vectors $\delta_{i}$ in the tight-binding model.
%(b) Spectrum of $H_\mathbf{k}$ with $t_1=.2,t_2=.3,t_3=.1$. Each band is doubly-degenerate. 
%(c) Spectrum of $H_\mathbf{k} + H^1_\mathbf{k} $ with $t_1=.2,t_2=.3,t_3=.1, t_4=.08,t_5=.05,t_6=.02$. 
%(d) Argument of the eigenvalues of the $z$-directed Wilson matrix along the path $\bar{M}-\bar{\Gamma}-\bar{M}'$ (blue dotted line in inset); $\bar{M} = (\pi,\pi)$, $\bar{\Gamma} = (0,0)$, $\bar{M}'=(-\pi,\pi)$.
(a) Spectrum of $H_\mathbf{k} + H^1_\mathbf{k} $ with $t_1=.2,t_2=.3,t_3=.1, t_4=.08,t_5=.05,t_6=.02$. Inset shows unit cell: orbitals on the blue atoms enter the Hamiltonian, while the white/black atoms create a crystal field with the symmetry of $P4_232$.
(b) Argument of the eigenvalues of the $z$-directed Wilson matrix along the path $\bar{M}-\bar{\Gamma}-\bar{M}'$ (blue dotted line in inset); $\bar{M} = (\pi,\pi)$, $\bar{\Gamma} = (0,0)$, $\bar{M}'=(-\pi,\pi)$.}
\end{figure}

We now consider the non-symmorphic simple cubic space group $P4_2 32$, which is generated by 
$\{ C_{2x} | \mathbf{0} \}$, $\{ C_{3,111} | \mathbf{0} \}$ and $\{ C_{2,110} | \frac{1}{2}\frac{1}{2}\frac{1}{2} \}$.
We also enforce time-reversal symmetry.
We consider atoms sitting at $(0,0,0)$ and $(\frac{1}{2},\frac{1}{2},\frac{1}{2})$ (inset to Fig~\ref{fig:208bands-nodeg}), which together comprise the $2a$ Wyckoff position, each with spinless $d_{z^2}$ and $d_{x^2-y^2}$ orbitals, which together form a time-reversal symmetric irrep %(denoted  $^1 \! E ^2 \! E$) 
of the site-symmetry group.\cite{ITA,PointGroupTables} % ($T$).
Since the orbitals transform as an irrep of a maximal Wyckoff position, any band structure derived from these orbitals transforms as a time-reversal invariant EBR.
It follows from Ref~\onlinecite{EBRTheory} that if the band structure is gapped, it contains topological bands.
%We will construct a simple model with a gapped band structure and directly prove the topological nature of the valence bands  by computing a Berry phase invariant.
Here, we explicitly construct a gapped Hamiltonian and a nontrivial bulk topological invariant, violating the conjecture\cite{Michel1999,Michel2001} that a single set of symmetry-related orbitals in a non-symmorphic space group always yields a gapless band structure.% in a time-reversal invariant system.

%Earlier work by Michel and Zak\cite{Michel1999,Michel2001} hypothesized that a single set of symmetry-related orbitals in a non-symmorphic space group would always yield a gapless band structure in a time-reversal invariant system.
%but our graph theory analysis in Refs~\onlinecite{GraphTheoryPaper,GroupTheoryPaper,GraphDataPaper} proved otherwise.

We consider the following Hamiltonian, which respects all space group symmetries and time-reversal:\footnote{\SUPPsymmetries\ shows the matrix form of the symmetries.}
\begin{align} 
%H_\mathbf{k} = H^1_\mathbf{k} + H^2_\mathbf{k} + H^3_\mathbf{k},
H_\mathbf{k} &= t_1 f_1(\mathbf{k})\tau_x \otimes \sigma_0 + t_2f_2(\mathbf{k}) \tau_y\otimes\sigma_0 +\nonumber\\
&+ t_3(g_1(\mathbf{k})\tau_z \otimes \sigma_z + g_2(\mathbf{k})\tau_z\otimes \sigma_x),
%H_\mathbf{k} &= (t_1 f_1(\mathbf{k})\tau_x + t_2f_2(\mathbf{k}) \tau_y)\sigma_0 + t_3\tau_z(g_1(\mathbf{k})  \sigma_z + g_2(\mathbf{k})\sigma_x),
\end{align}
where
%\begin{align}
%f_1(\mathbf{k})&=\cos((k_x+k_y+k_z)/2)+\cos((-k_x-k_y+k_z)/2)\nonumber\\
%&+\cos((-k_x+k_y-k_z)/2)+\cos((k_x-k_y-k_z)/2) \nonumber\\
%f_2(\mathbf{k})&=\sin((k_x+k_y+k_z)/2)+\sin((-k_x-k_y+k_z)/2)\nonumber\\
%&+\sin((-k_x+k_y-k_z)/2)+\sin((k_x-k_y-k_z)/2),\nonumber\\
%g_1(\mathbf{k}) &= \cos k_x - \cos k_y\nonumber\\
%g_2(\mathbf{k}) &= \frac{1}{\sqrt{3}}\left(\cos k_x+\cos k_y - 2\cos k_z\right)
%\end{align}
$f_1(\mathbf{k}) \!=\! \sum_i \cos (\mathbf{k}\cdot \delta_i)$, 
$f_2(\mathbf{k}) \!=\! \sum_i \sin (\mathbf{k}\cdot \delta_i)$, 
$g_1(\mathbf{k}) = \cos k_x - \cos k_y$, 
$g_2(\mathbf{k}) = \left(\cos k_x+\cos k_y - 2\cos k_z\right)/\sqrt{3} $
and $\delta_{1,2,3,4}$ are vectors to nearest neighbors, shown in Fig~\ref{fig:208bands-nodeg}.
The band structure is doubly-degenerate and gapped when $t_{1,2,3}\neq 0$.
To eliminate the extra degeneracies, we add the symmetry-preserving term
%besides those required at $\Gamma = (0,0,0)$, $R=(\pi,\pi,\pi)$, $X=(\pi,0,0)$, and $M=(\pi,\pi,0)$.
\begin{align}
H_\mathbf{k}^1 &= t_4f_1(\mathbf{k}) \tau_y\otimes \sigma_y + t_5\tau_0\otimes (g_2(\mathbf{k}) \sigma_z -g_1(\mathbf{k})  \sigma_x ) \nonumber\\
&+ t_6 g_3(\mathbf{k}) \tau_z \otimes \sigma_0,
\end{align}
where
\begin{align}
g_3(\mathbf{k}) &= \cos(2k_x)\cos(k_y) \!-\! \cos(2k_y)\cos(k_x) + \text{perm.}
%\cos(2k_x)\cos(k_y)+\cos(2k_y)\cos(k_z)\nonumber\\
%&+\cos(2k_z)\cos(k_x) - \cos(2k_y)\cos(k_x)\nonumber\\
%&-\cos(2k_z)\cos(k_y)-\cos(2k_x)\cos(k_z) 
\end{align}
and `+ perm' indicates terms obtained by permuting $k_x\rightarrow k_y\rightarrow k_z$.
The spectrum of $H_\mathbf{k} + H_\mathbf{k}^1$ is shown in Fig~\ref{fig:208bands-nodeg}.
%The only required degeneracies are at the high symmetry points.
Since $H_\mathbf{k}$ is fully gapped when $t_{1,2,3}\neq 0$, $H_\mathbf{k} + H_\mathbf{k}^1$ is gapped when $t_{4,5,6}$ are small compared to $t_{1,2,3}$.
%For all parameter values, $H_\mathbf{k} + H_\mathbf{k}^1$ commutes with the symmetries in \SUPPsymmetries.

%\paragraph{Wilson Hamiltonian}

%We argued in the previous section that the Hamiltonian $H_\mathbf{k} + H_\mathbf{k}^1$ is gapped when $t_{4,5,6} \ll t_{1,2,3}$.
This gapped phase realizes a disconnected and time-reversal symmetric EBR; thus, it contains topological bands.
We diagnose the topological phase by the winding of its $z$-directed Wilson loop along the \emph{bent} path shown in Fig~\ref{fig:208wilsonz}.
This is a time-reversal symmetric and non-symmorphic generalization of the ``bent Chern number'' introduced in Ref~\onlinecite{Alexandradinata14a}.
Two features are necessary for this loop to wind:
first, the Wilson loop eigenvalues are pinned to $\pm 1$ at $\bar{\Gamma}$ and $\bar{M}$,
and, second, there are protected band crossings in the Wilson spectrum along the $|k_x| = |k_y|$ lines.
Combined, these features prevent the Wilson bands from being smoothly deformable to flat bands; hence, the phase is topological.
%The first observation is explained because the $C_{2x}(C_{2y})$ symmetry along $\bar{\Gamma}-\bar{X}(\bar{X}-\bar{M})$ forces the eigenvalues of $\mathcal{W}_\mathbf{k}$ to either be real or come in complex conjugate pairs; 

We now explain the origin of these features: first, $C_{2x}$ symmetry forces the eigenvalues of $\mathcal{W}_{(\bar{\Gamma},0)}$ and $\mathcal{W}_{(\bar{M},0)}$ to be real, while the $\{ C_{2,110}| \frac{1}{2}\frac{1}{2}\frac{1}{2}\}$ screw symmetry forces them to come in pairs $(\lambda, -\lambda^*)$.\footnote{See \SUPPwilsongapped .}
This combination pins the eigenvalues of $\mathcal{W}_{(k_x,k_y,0)}$ to be $\pm 1$ at $\bar{\Gamma}$ and $\bar{M}$.
% (as well as along the whole path $\bar{\Gamma}-\bar{X} - \bar{M}-\bar{Y}-\bar{\Gamma}$).

The Wilson band crossing is subtle: the $\{ C_{2,110}| \frac{1}{2}\frac{1}{2}\frac{1}{2}\}$ screw symmetry requires the eigenvalues of $\mathcal{W}_{(k,k,0)}$ to come in pairs $(\lambda, -\lambda^*)$.
Combined with the anti-unitary symmetry $\mathcal{T}\{ C_{2,110}| \frac{1}{2}\frac{1}{2}\frac{1}{2}\}^{-1}C_{2z}$, which leaves points $(k,k,k_z)$ invariant, the Wilson matrix must take the form $\mathcal{W}_{(k,k,0)} = ie^{ia_x(k)\sigma_x + ia_y(k)\sigma_y}$, where, importantly, $a_x(k) \propto a_y(k)$ (alternately, the symmetries permit the eigenvalues of $\mathcal{W}_{(k,k,0)}$ to be fixed to $\pm 1$; see \SUPPwilsoncrossing.)
Then degeneracies of the Wilson eigenvalues, which occur when $a_x(k)=a_y(k)=0$, are not fine-tuned, since the symmetry forced $a_x(k)\propto a_y(k)$. Since the eigenvalues of $\mathcal{W}_{(k,k,0)}$ at $k=0$ and $k=\pi$ are fixed to $+1$ and $-1$, an odd number of linear degeneracies between $\bar{\Gamma}$ and $\bar{M}$ cannot be removed without closing the bulk band gap.
Thus, the parity of the number of linear degeneracies is a topological invariant.
%An odd number of band crossings implies that the phase of each Wilson band winds at least once from $0$ to $2\pi$ along the closed path $\bar{M}-\bar{\Gamma}-\bar{M}'$, %($\bar{M}=(\pi,\pi)$ and $\bar{M}'=(-\pi,\pi)$ differ by a reciprocal lattice vector), 
%as shown in Fig~\ref{fig:208wilsonz}.\footnote{See \SUPPbentwinding .}
%The winding indicates the valence bands are topological.
%The winding occurs because the product of $C_{2x}$ and the screw symmetry ($\{C_{2,110}|\frac{1}{2}\frac{1}{2}\frac{1}{2} \}$) relates the Wilson loop eigenvalues along $\bar{\Gamma}-\bar{M}$ to $\bar{\Gamma}-\bar{M}'$ with a minus sign (see Appendix~\ref{sec:208wilsonsymm}).

%Since the Wilson bands are non-degenerate along the path $\gamma = \bar{\Gamma}-\bar{X}-\bar{M}-\bar{Y}-\bar{\Gamma}$ that surrounds the band crossing, it represents
The band crossing forms a Dirac cone in the two-dimensional ``Wilson Hamiltonian.''\cite{ArisCohomology}
The Dirac point %(and thus the nontrivial nature of the insulating phase) 
is revealed by the Berry phase, $w$, acquired by an eigenstate of $\mathcal{W}_{(k_x,k_y,0)}$ as it traverses the path $\gamma$ around the Dirac point.
The Berry phase of Wilson loop eigenstates was introduced in Ref.~\onlinecite{hughesbernevig2016}.
Since $w$ is quantized to $\pm 1$ (see \SUPPwilsofwils), it constitutes a topological invariant.
%Furthermore, since the Wilson bands of the trivial phase are flat, the path $\gamma$ is contractible, which proves that in the trivial phase, $w=1$.
In our model, for several values of parameters, we have numerically computed the nontrivial value, $w=-1$.

When SOC is present, the spinful $d_{z^2}$ and $d_{x^2-y^2}$ orbitals transform as spin-$\frac{3}{2}$ orbitals, which induce an eight-band time-reversal symmetric EBR.\cite{GroupTheoryPaper}
When the EBR is gapped, the valence (or conduction) bands must be topological.

\paragraph{Weak symmetry indicators}

In both the spinless TCI on the honeycomb lattice and the non-symmorphic gapped EBR,
the valence bands are topological, but have the property that the irreps at high-symmetry points can be written as a ``difference'' of the irreps in two other EBRs.\footnote{Details for the honeycomb case are in \SUPPpxyirreps .}
Because the irreps can be written as a difference, classification schemes\cite{Po2017} that treat the little group irreps as a vector space will identify the valence bands as trivial, even though they lack an atomic limit.
However, unless an energy gap closes to the valence bands, the winding of the Wilson loop in both examples provides a robust and quantized topological invariant that is, in principle, physically observable.\cite{Atala13,Nakajima16,Holler2017}

This distinction warrants a refined characterization of topological crystalline bands based on whether their topological nature can be deduced by their little group irreps.
We label the symmetry properties of topological bands as strong if their little group irreps are not equal to a linear combination of little group irreps corresponding to EBRs and weak if their little group irreps are equal to a difference (but not a sum) of irreps in EBRs.
Strong symmetry properties implies a stable topological index; however, the converse is not true: for example, bands with a nontrivial $\mathbb{Z}_2$ index under time-reversal symmetry can be strong\cite{Fu2007} or weak.\cite{NaturePaper}
This usage of weak and strong symmetry is different than the current distinction between weak and strong topological insulators.\cite{Fu2007}
It is more suitable for the refined classification of topological insulators with crystal symmetries.

%%%%%%%%%%%%%%%%%%%%%%%%%%%%%%
%%%%%%%%%%%%%%%%%%%%%%%%%%%%%%
%%%%%%%%%%%%%%%%%%%%%%%%%%%%%%
%%%%%%%%%%%%%%%%%%%%%%%%%%%%%%
\paragraph{Conclusions}

We have constructed tight-binding models to realize the insulating phases of two gapped EBRs. 
We explicitly showed that the valence bands have a nontrivial topological invariant.
In doing so, we found a new topological invariant in a non-symmorphic space group: a Dirac cone in the Wilson loop spectrum and a Wilson loop that winds along a bent path.
This motivates further study of the gapped EBRs in other non-symmorphic space groups.
In addition, we introduced the notion of a weak symmetry indicator.
We postpone a general investigation of the symmetry properties of gapped EBRs to future work.

\begin{acknowledgments}
\paragraph{Acknowledgments}
BB, JC, ZW, and BAB acknowledge the hospitality of the Donostia International Physics Center.
BAB acknowledges the hospitality and support of the \'{E}cole Normale Sup\'{e}rieure and Laboratoire de Physique Th\'{e}orique et Hautes Energies. 
The work of MGV was supported by FIS2016-75862-P.
The work of LE and MIA was supported by the Government of the Basque Country (project IT779-13)  and the Spanish Ministry of Economy and Competitiveness and FEDER funds (project MAT2015-66441-P). 
BAB acknowledges the support of the NSF EAGER Award DMR -- 1643312, ONR - N00014-14-1-0330, NSF-MRSEC DMR-1420541, ARO MURI W911NF-12-1-0461, the Department of Energy de-sc0016239, the Simons Investigator Award, Packard Foundation and the Schmidt Fund for Innovative Research.

\end{acknowledgments}

\bibliography{NewTIs}
\end{document}

\title{Supplemental Material for Topology of Disconnected Elementary Band Representations}
\author{Jennifer Cano}
\affiliation{Princeton Center for Theoretical Science, Princeton University, Princeton, New Jersey 08544, USA}
\author{Barry Bradlyn}
\affiliation{Princeton Center for Theoretical Science, Princeton University, Princeton, New Jersey 08544, USA}
\author{Zhijun Wang}
\affiliation{Department of Physics, Princeton University, Princeton, New Jersey 08544, USA}
\author{L. Elcoro}
\affiliation{Department of Condensed Matter Physics, University of the Basque Country UPV/EHU, Apartado 644, 48080 Bilbao, Spain}
\author{M.~G. Vergniory}
\affiliation{Donostia International Physics Center, P. Manuel de Lardizabal 4, 20018 Donostia-San Sebasti\'{a}n, Spain}
\affiliation{Department of Applied Physics II, University of the Basque Country UPV/EHU, Apartado 644, 48080 Bilbao, Spain}
\affiliation{Ikerbasque, Basque Foundation for Science, 48013 Bilbao, Spain}
%\affiliation{Max Planck Institute for Solid State Research, Heisenbergstr. 1,
%70569 Stuttgart, Germany.}
\author{C. Felser}
\affiliation{Max Planck Institute for Chemical Physics of Solids, 01187 Dresden, Germany}
\author{M.~I.~Aroyo}
\affiliation{Department of Condensed Matter Physics, University of the Basque Country UPV/EHU, Apartado 644, 48080 Bilbao, Spain}
\author{B. Andrei Bernevig}
\thanks{Permanent Address: Department of Physics, Princeton University, Princeton, New Jersey 08544, USA }
\affiliation{Department of Physics, Princeton University, Princeton, New Jersey 08544, USA}
\affiliation{Donostia International Physics Center, P. Manuel de Lardizabal 4, 20018 Donostia-San Sebasti\'{a}n, Spain}
\affiliation{Laboratoire Pierre Aigrain, Ecole Normale Sup\'{e}rieure-PSL Research University, CNRS, Universit\'{e} Pierre et Marie Curie-Sorbonne Universit\'{e}s, Universit\'{e} Paris Diderot-Sorbonne Paris Cit\'{e}, 24 rue Lhomond, 75231 Paris Cedex 05, France}
\affiliation{Sorbonne Universit\'{e}s, UPMC Univ Paris 06, UMR 7589, LPTHE, F-75005, Paris, France}
\affiliation{LPTMS, CNRS (UMR 8626), Universit\'e Paris-Saclay, 15 rue Georges Cl\'emenceau,\\ 91405 Orsay, France}

\maketitle

\section{Nearest neighbor Hamiltonian for $p_{x,y}$ orbitals on the honeycomb lattice}
\label{sec:pxyHam}

We choose the lattice basis vectors:
\begin{align}
\mathbf{e}_1 &= \frac{\sqrt{3}}{2}\hat{\mathbf{x}} + \frac{1}{2} \hat{\mathbf{y}} \nonumber\\
\mathbf{e}_2 &= \frac{\sqrt{3}}{2}\hat{\mathbf{x}} - \frac{1}{2} \hat{\mathbf{y}},
\end{align}
which are shown in \MAINfiggraphenebasisvectors\ with
their reciprocal lattice vectors, $\mathbf{g}_i$, which satisfy $\mathbf{g}_i\cdot \mathbf{e}_j = 2\pi \delta_{ij}$.
Sites on the $A(B)$ sublattice sit at positions $\mathbf{R} + \mathbf{r}_{A(B)}$, where $\mathbf{R}$ denotes a lattice translation and 
\begin{align}
\mathbf{r}_A &= \frac{2}{3}\mathbf{e}_1 - \frac{1}{3}\mathbf{e}_2\nonumber\\
\mathbf{r}_B &= \frac{1}{3}\mathbf{e}_1 - \frac{2}{3}\mathbf{e}_2
\end{align}
In this basis, the symmetry generators of the honeycomb lattice act as follows:
\begin{align}
C_{3z} &:  (\mathbf{e}_1 , \mathbf{e}_2) \rightarrow (-\mathbf{e}_2, \mathbf{e}_1 - \mathbf{e}_2) \nonumber\\
C_{2z} &:  (\mathbf{e}_1 , \mathbf{e}_2) \rightarrow (-\mathbf{e}_1 , -\mathbf{e}_2) \nonumber\\
m_{1\bar{1}} &: (\mathbf{e}_1 , \mathbf{e}_2) \rightarrow (\mathbf{e}_2 , \mathbf{e}_1),
\end{align}  
where the subscript $1\bar{1}$ indicates that the mirror plane has normal vector $\mathbf{e}_1 - \mathbf{e}_2$.
For $p_{x,y}$ orbitals, we choose the following matrix representation, in which the Pauli matrices $\tau$ act in sublattice space and the $\sigma$ matrices act in orbital space:
\begin{align}
U_{C_{3z}} &=  \tau_0 \otimes \left(-\frac{1}{2}\sigma_0 + i \frac{\sqrt{3}}{2}\sigma_y\right) \nonumber\\
U_{C_{2z}} &= -\tau_x \otimes \sigma_0 \nonumber\\
U_{m_{1\bar{1}}} &= \tau_0 \otimes \sigma_z,
\label{eq:unitarygensxy}
\end{align}
The orbital term for a rotation by an angle $\theta$ about an axis $\hat{\mathbf{n}}$ is represented by $e^{i\theta \hat{\mathbf{n}} \cdot \mathbf{S}}$, projected onto the $p_{x,y}$ orbitals; $\mathbf{S}$ is the vector of spin-1 matrices.
A Hamiltonian, $H_\mathbf{k}$, that respects the lattice symmetry must satisfy:
\begin{equation}
H_\mathbf{k} = U_R^\dagger H_{R\mathbf{k}} U_R,
\end{equation} 
for each generator, $R$, of the honeycomb lattice.

Denoting the annihilation operator on site $\mathbf{r}$ by $c_{\mathbf{r},a}$, where $a = x,y$ indicates the $p_x$ or $p_y$ orbital, the nearest neighbor Hamiltonian is given by:
\begin{equation}
H = \sum_{\mathbf{R}}\sum_{a,b} \sum_{\mathbf{\delta}_i}    t_{ab}(\delta_i)    c_{\mathbf{R} +\mathbf{r}_A, a}^\dagger  c_{\mathbf{R} + \mathbf{r}_A+ \delta_i, b}  + {\rm h.c.},
\label{eq:pxyHamreal}
\end{equation}
where the three nearest neighbors to a site at $\mathbf{R}+\mathbf{r}_A$ sit at $\mathbf{R}+\mathbf{r}_A + \delta_i$ (the vectors $\delta_i$ are depicted in \MAINfiggraphenebasisvectors) and
%\begin{align}
%\delta_1 &= -\frac{1}{3}(\mathbf{e}_1 + \mathbf{e}_2) \nonumber\\
%\delta_2 &= \frac{1}{3}(2\mathbf{e}_1 - \mathbf{e}_2)\nonumber\\
%\delta_3 &= \frac{1}{3}(-\mathbf{e}_1 + 2\mathbf{e}_2)
%\label{eq:defdelta}
%\end{align}
$t_{ab}(\delta_i)$ is given by one of the Slater-Koster terms:\cite{SlaterKoster}
\begin{align}
%t_{xx}( \delta_i ) &= \delta_{ix}^2t_\sigma + (1-\delta_{ix}^2)t_\pi \nonumber\\
%t_{yy}( \delta_i ) &= \delta_{iy}^2t_\sigma + (1-\delta_{iy}^2)t_\pi \nonumber\\
%t_{xy}( \delta_i ) &= \delta_{ix}\delta_{iy} (t_\sigma - t_\pi) = t_{yx}( \delta_i),
t_{xx}( \delta_i ) &= \frac{1}{3}\left[ (\delta_i \cdot \hat{\mathbf{x}})^2  t_\sigma + (\delta_i \cdot \hat{\mathbf{y}})^2 t_\pi \right] \nonumber\\
t_{yy}( \delta_i ) &=  \frac{1}{3} \left[ (\delta_i \cdot \hat{\mathbf{y}})^2  t_\sigma + (\delta_i \cdot \hat{\mathbf{x}})^2 t_\pi \right] \nonumber\\
t_{xy}( \delta_i ) &= \frac{1}{3} (\delta_i \cdot \hat{\mathbf{x}})(\delta_i \cdot \hat{\mathbf{y}})(t_\sigma - t_\pi) = t_{yx}( \delta_i),
\label{eq:hopping}
\end{align}
where $t_{\sigma(\pi)}$ are free parameters that describe $\sigma(\pi)$-bonds.
Notice that $\mathbf{r}_A+\delta_i$ is always a site on the $B$ sublattice.
Using the Fourier transform,
\begin{equation}
c_{\mathbf{k},L,a} = \sum_\mathbf{R} e^{i\mathbf{k} \cdot (\mathbf{R}+\mathbf{r}_{L}) }c_{\mathbf{R} + \mathbf{r}_L,a}
\end{equation}
where $L=A,B$ denotes the sublattice and $a=x,y$ denotes the orbital,
the real space Hamiltonian in Eq~(\ref{eq:pxyHamreal}) is rewritten:
\begin{equation}
H = \sum_\mathbf{k} \sum_{a,b}\sum_{\delta_i} c_{\mathbf{k},A,a}^\dagger c_{\mathbf{k},B,b}e^{-i\mathbf{k}\cdot \delta_i} t_{ab}(\delta_i) + {\rm h.c.},
\label{eq:pxyHam1}
\end{equation}
% and $\delta_{i} = |\delta_i|\left( \delta_{ix}\hat{\mathbf{x}}+\delta_{iy}\hat{\mathbf{y}} \right)$. 
Plugging Eq~(\ref{eq:hopping}) into Eq~(\ref{eq:pxyHam1}) yields 
\begin{equation}
H^0 =\sum_\mathbf{k} \psi_\mathbf{k}^\dagger H_\mathbf{k}^0 \psi_\mathbf{k} \equiv \sum_\mathbf{k} \psi_\mathbf{k}^\dagger \begin{pmatrix} 0 & h_\mathbf{k} \\ h_\mathbf{k}^\dagger & 0 \end{pmatrix} \psi_\mathbf{k}
\end{equation}
where $\psi_\mathbf{k} = \begin{pmatrix} c_{\mathbf{k},A,x} & c_{\mathbf{k},A,y} & c_{\mathbf{k},B,x} & c_{\mathbf{k},B,y} \end{pmatrix}^T$ contains the annihilation operators for $p_{x,y}$ orbitals on the $A$ and $B$ sublattices and $h_\mathbf{k}$ is given by \MAINeqpxyHamh .

\subsection{Phase diagram of $H_\mathbf{k}^0 + xH_\mathbf{k}^1$}
\label{sec:tciphase}

As described in the main text,
%Sec~\ref{sec:spinlessTCI}, 
a necessary condition to reach the gapped TCI phase is that the two-fold degeneracy at $K$ is higher or lower in energy than the other two bands.
The eigenvalues of $H_{\mathbf{k}=K}^0 + x H_{\mathbf{k}=K}^1$ are $-\frac{3\sqrt{3}}{8}x$ (2-fold degenerate) and $\frac{3}{8}(\pm 4(t_\pi - t_\sigma)+\sqrt{3}x )$; 
thus, the TCI phase requires: 
\begin{equation}
|x|> \frac{2}{\sqrt{3}}|t_\pi - t_\sigma| 
\label{eq:TCIcond}
\end{equation}
%$x> \frac{2}{\sqrt{3}}|t_\pi - t_\sigma| $ or $x < -\frac{2}{\sqrt{3}}|t_\pi - t_\sigma|$.

However, there are further constraints: band crossings along the paths connecting $\Gamma$ and $M$ can prevent the system from opening a gap even when the energy ordering at $K$ allows it.
Consider the two lines $\Sigma = \alpha \mathbf{g}_1$ and $\Lambda = \beta(\frac{1}{3}\mathbf{g}_1 + \frac{2}{3} \mathbf{g}_2)$, which are invariant under $C_{3z}m_{1\bar{1}}$ and $C_{2z}C_{3z}m_{1\bar{1}}$, respectively.
As $\alpha$ goes from $0$ to $\frac{1}{2}$, $\Sigma$ connects $\Gamma$ to $M$.
As $\beta$ goes from $0$ to $\frac{3}{2}$, $\Lambda$ connects $\Gamma$ to $M + \mathbf{g}_2$ (passing through $K'$ at $\beta = 1$).
We can then track the respective mirror eigenvalues along these lines to find constraints on connectivity.
The BANDREP application on the BCS server shows that the $\Gamma_5$ and $\Gamma_6$ irreps appear at $\Gamma$ in our model ($p_{x,y}$ orbitals on the honeycomb correspond to the irrep $E$ at Wyckoff position $2b$).
In the $\Gamma_5$ and $\Gamma_6$ irreps, the character of each mirror is zero: this means that the doubly-degenerate bands that comprise the $\Gamma_5$ irrep will split into two bands along $\Sigma$, one of which is even under $C_{3z}m_{1\bar{1}}$ and one of which is odd (and same for $\Gamma_6$).
Thus, $\Gamma_5$ must connect to two irreps at $M$ which have opposite parity under $C_{3z}m_{1\bar{1}}$.
Similarly, the $\Gamma_5$ irrep will split into two bands along $\Lambda$, one of which is even under $C_{2z}C_{3z}m_{1\bar{1}}$ and one of which is odd; thus, $\Gamma_5$ must connect to two irreps at $M$ which have opposite parity under $C_{2z}C_{3z}m_{1\bar{1}}$.
Without loss of generality, assume the $\Gamma_5$ irrep is higher in energy than $\Gamma_6$; 
then in order for the system to be an insulator, 
the two bands that are highest in energy at $M$ must have opposite $C_{3z}m_{1\bar{1}}$ eigenvalues and also opposite $C_{2z}C_{3z}m_{1\bar{1}}$ eigenvalues.

\begin{figure}[h]
\centering
\subfloat[]{
	\includegraphics[width=1.5in]{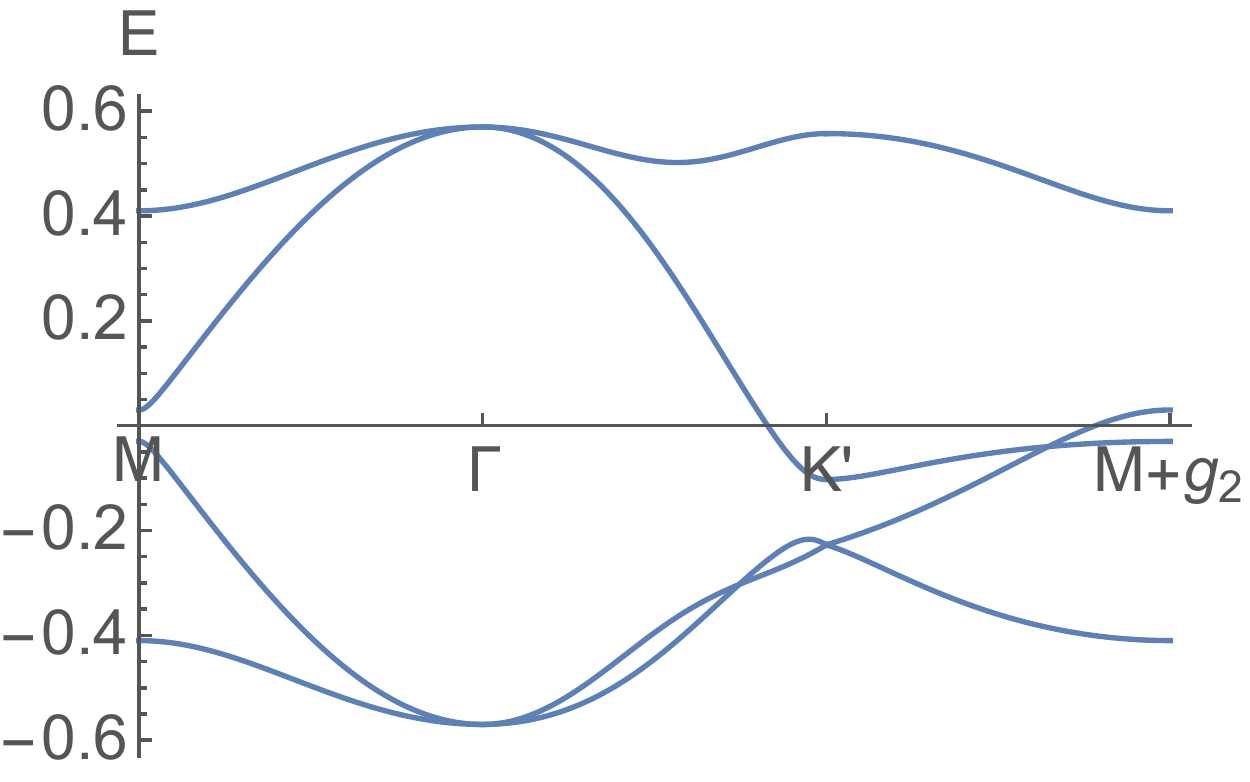}\label{fig:Weyl}
	\begin{picture}(0,0)
		\put(-30,35){\includegraphics[height=1.6cm]{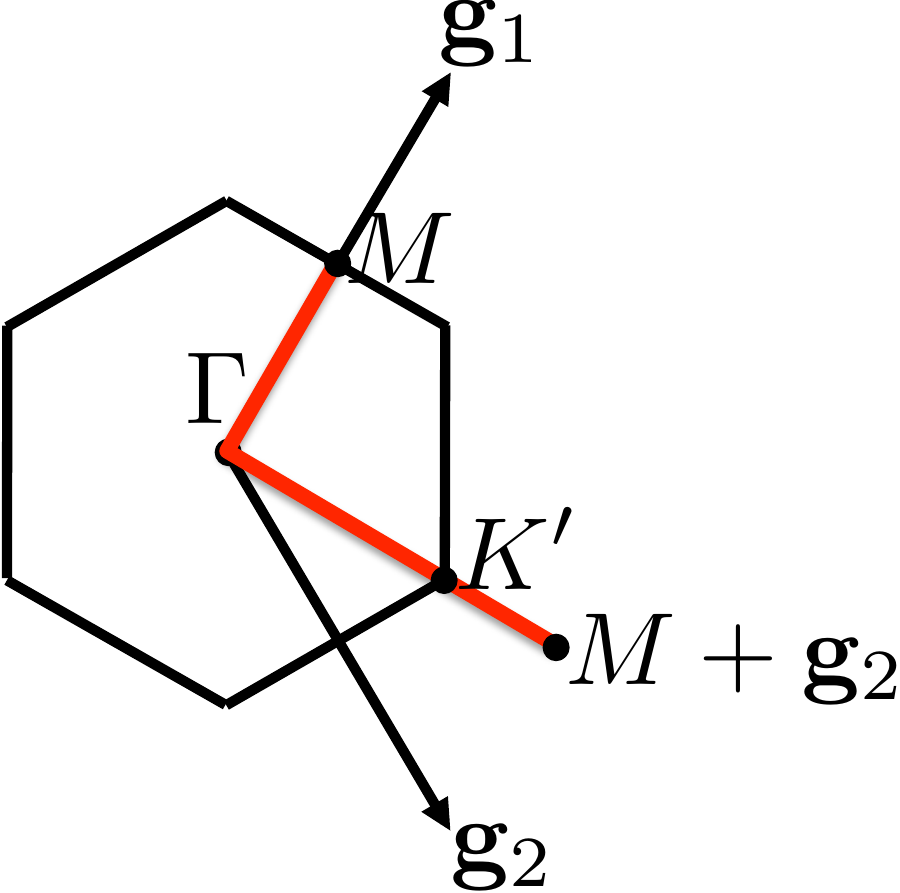}}
	\end{picture}
	}\quad
\subfloat[]{
	\includegraphics[width=1.5in]{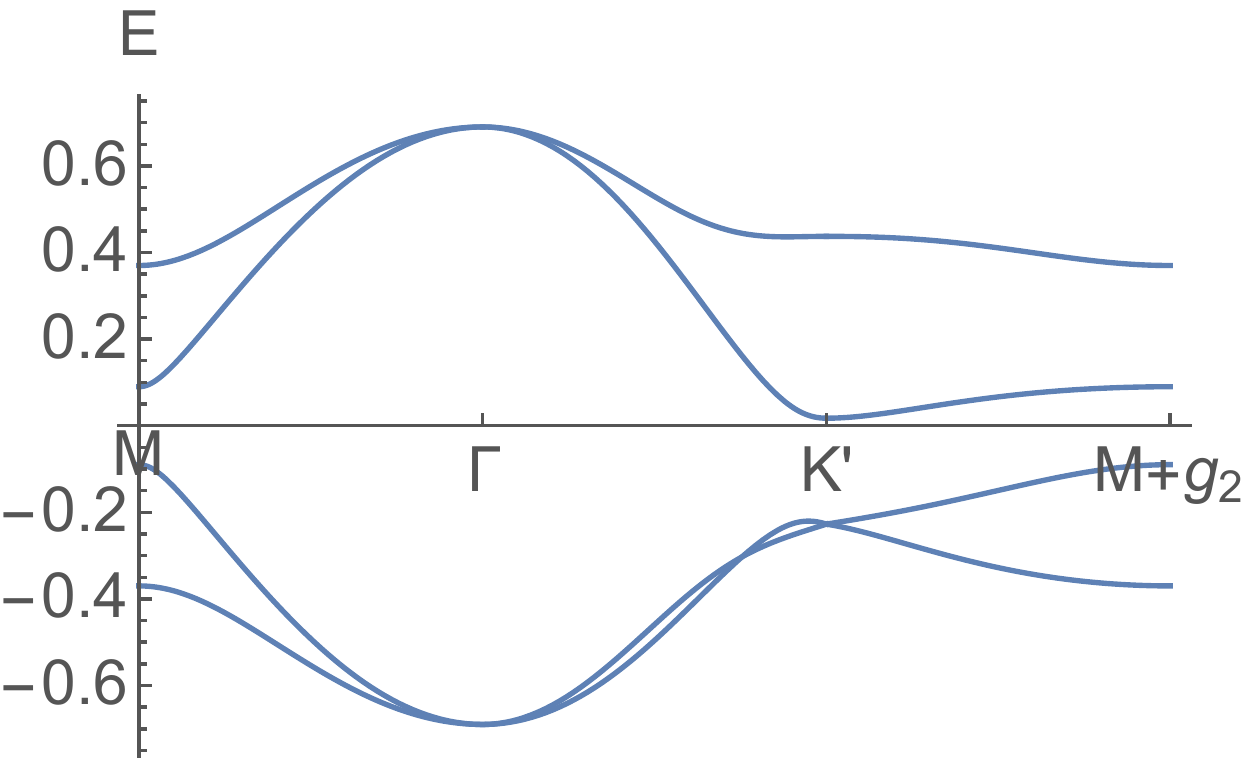}\label{fig:Weylgone}
}
\caption{Band structure of $H^0_\mathbf{k}+xH^1_\mathbf{k}$ along the red path shown in the inset; $x=.35$. In (a) $t_\sigma=.08, t_\pi = .3$, while in (b) $t_\sigma = .16, t_\pi = .3$.
In both cases, Eq~(\ref{eq:TCIcond}) is satisfied; hence the band ordering at $K$ allows for a gap.
However, since the parameters in (a) violate Eq~(\ref{eq:pxygap1}) and (\ref{eq:pxygap2}), there is a band crossing between $\Gamma$ and $M+\mathbf{g}_2$. In (b), Eq~(\ref{eq:pxygap2}) is satisfied and the band structure is gapped.}
\end{figure}

%The quickest way to deduce this information is using the theory of topological quantum chemistry as implemented on the Bilbao Crystallographic Server (BCS).\cite{GroupTheoryPaper,NaturePaper,EBRTheory}
%Inputting $p_{x,y}$ orbitals at the honeycomb corners (the irrep $E$ at the Wyckoff position $2b$) in space group $P6mm$ (183) into the BANDREP application shows that two two-dimensional irreps, $\Gamma_5 \oplus \Gamma_6$, will appear in the band structure.
%Then consider the two lines connecting $\Gamma$ and $M$, denoted $\Sigma = \alpha \mathbf{g}_1$ and $\Lambda = \beta(\frac{1}{3}\mathbf{g}_1 + \frac{2}{3} \mathbf{g}_2)$, which are invariant under $C_{3z}m_{1\bar{1}}$ and $C_{2z}C_{3z}m_{1\bar{1}}$, respectively.
%The COMPREL application shows that $\Gamma_5$ splits into $\Sigma_1 \oplus \Sigma_2$; $\Sigma_{1}(\Sigma_2)$ is even(odd) under $C_{3z}m_{1\bar{1}}$; similarly, along the other line, $\Gamma_5$ splits into the even(odd) irreps $\Lambda_1(\Lambda_2)$.
%In order for the band structure to be disconnected, the $\Gamma_5$ irrep must connect to the same two irreps at the $M$ point, whether following the path $\Sigma$ or $\Lambda$.
%Thus, if $\Gamma_5$ is higher(lower) in energy than $\Gamma_6$, a gapped band structure requires that the two irreps at $M$ with the highest(lowest) energies must have opposite eigenvalues under both $C_{3z}m_{1\bar{1}}$ and opposite eigenvalues under $C_{2z}C_{3z}m_{1\bar{1}}$. 
%Using the eigenvalues in Table~\ref{tab:Meigs}, 
Referring to Table~\ref{tab:Meigs}, a gapped band structure requires one of the two following conditions to be satisfied:
\begin{align}
 E_{M1,M2} > E_{M3,M4} \Rightarrow 3t_\sigma < t_\pi < \frac{1}{3}t_\sigma \label{eq:pxygap1}\\
 E_{M1,M2} < E_{M3,M4} \Rightarrow \frac{1}{3}t_\pi < t_\sigma < 3t_\pi  \label{eq:pxygap2}
\end{align}
If neither Eq~(\ref{eq:pxygap1}) or (\ref{eq:pxygap2}) is satisfied, there will be a band crossing along $\Gamma-M$, as shown in Fig~\ref{fig:Weyl}.

The $C_{2z}$ eigenvalues at $\Gamma$ are shown in Table~\ref{tab:Geigs}.
By comparing Table~\ref{tab:Meigs} with Table~\ref{tab:Geigs}, one can check that in the gapped phase, when either Eq~(\ref{eq:pxygap1}) or (\ref{eq:pxygap2}) is satisfied, the $C_{2z}$ eigenvalues at $\Gamma$ are always opposite those at $M$ (unlike the usual time-reversal $\mathbb{Z}_2$ invariant, here we are referring to the $C_{2z}$ eigenvalues themselves, not their product. %(per the discussion in Sec~\ref{sec:spinlessTCI} and the invariant defined in Ref~\onlinecite{Alexandradinata14}).
This guarantees that the Wilson loop in \MAINfigpxyWilson\ always winds in the gapped phase.\cite{Alexandradinata14}
%, as proved in  Sec~\ref{sec:spinlessTCI}.
%The constraints are independent of $x$ because $H^1_{\mathbf{k}=M}=0$.

We note that it could be possible with longer range hopping terms to reach a gapped phase where the $C_{2z}$ eigenvalues of the two lower bands at $\Gamma$ are the same as those at $M$ (this is symmetry-allowed according to possible decompositions of the EBR induced from $p_{x,y}$ orbitals listed on the BCS Server\cite{GroupTheoryPaper}). However, this phase is not accessible within our nearest-neighbor model.

\begin{table}[h]
{\renewcommand{\arraystretch}{1.2}%
\centering
\begin{tabular}{c|c|c|c}
Energy at $M$ & $C_{3z}m_{1\bar{1}}$ eig. & $C_{2z}C_{3z}m_{1\bar{1}}$ eig. & $C_{2z}$ eig.\\
\hline
$E_{M1}\equiv \frac{t_\pi}{2} - \frac{3t_\sigma}{2}$ & $-1$ & $1$ & $-1$ \\
$E_{M2}\equiv -\frac{3t_\pi}{2} + \frac{t_\sigma}{2}$ &$1$ & $-1$ & $-1$\\
$E_{M3} \equiv \frac{3t_\pi}{2} - \frac{t_\sigma}{2}$ & $1$ & $1$ & $+1$ \\
$E_{M4} \equiv -\frac{t_\pi}{2} + \frac{3t_\sigma}{2}$ & $-1$ & $-1$ & $+1$
\end{tabular}}
\caption{Energies, mirror, and $C_{2z}$ eigenvalues for each of the eigenstates at $M$. The $C_{2z}$ eigenvalue is a product of the two mirror eigenvalues.}
\label{tab:Meigs}
\end{table}

\begin{table}[h]
{\renewcommand{\arraystretch}{1.2}%
\begin{tabular}{c|c}
Energy at $\Gamma$ & $C_{2z}$ eig.\\
\hline 
$E_{\Gamma 1}= -\frac{3}{2}(t_\pi + t_\sigma)$ & $+1, +1$\\
$E_{\Gamma 2}= \frac{3}{2}(t_\pi + t_\sigma)$ & $-1, -1$
\end{tabular}}
\caption{Energies and $C_{2z}$ eigenvalues for the two-fold degenerate eigenstates at $\Gamma$.}
\label{tab:Geigs}
\end{table}

\subsection{Irreps at high-symmetry points}
\label{sec:pxyirreps}

Using the notation on BANDREP application of the BCS\cite{GroupTheoryPaper}, the Hamiltonian $H_\mathbf{k}^0 + xH_\mathbf{k}^1$ can realize two possible sets of valence bands: $(\Gamma_5, K_3, M_3, M_4)$ or $(\Gamma_6, K_3, M_1, M_2)$, assuming, without loss of generality, that the $K_3$ irrep appears in the valence bands instead of the conduction bands.
(As noted at the end of the last section, there are two other disconnected phases listed in the BANDREP application that that our model does not realize and which differ in the $C_{2z}$ eigenvalues of occupied bands; presumably they require longer range hopping).
By comparing to the list of band representations induced from Wyckoff positions in $P6mm$ (SG 183) (which describes layers of the honeycomb lattice with no additional symmetry in the $z$ direction), we see that the irreps in $A_1\uparrow G$ or $B_2\uparrow G$ from the $1a$ position are $(\Gamma_1, K_1, M_1)$ or $(\Gamma_3, K_1, M_3)$, respectively, while the irreps in $A_1\uparrow G$ on the $3c$ position are $(\Gamma_1, \Gamma_5, K_1, K_3, M_1, M_3, M_4)$ and the irreps in $B_1\uparrow G$ on the $3c$ position are $(\Gamma_3, \Gamma_6, K_1, K_3, M_1, M_2, M_3)$.
Thus, we see that for each of the possible sets of valence bands in our model, 
the irreps that appear are obtained from subtracting the irreps of one of the EBRs induced from the $1a$ position from one of the EBRs induced from the $3c$ position.
A classification scheme that only looks at the irreps at high-symmetry points will classify our valence bands as trivial.
It was noted in Ref~\onlinecite{Po2017} that some topologically nontrivial bands will be included in the trivial class; our model is an example of this phenomenon.

\subsection{Phase diagram with SOC}
\label{sec:SOCxy}

Including SOC, 
$\Gamma_{5} \rightarrow \bar{\Gamma}_7 \oplus \bar{\Gamma}_{8}$, $\Gamma_6\rightarrow \bar{\Gamma}_7 \oplus \bar{\Gamma}_{9}$
and
$K_{1} \rightarrow \bar{K}_6$, $K_{2} \rightarrow \bar{K}_6$, $K_3 \rightarrow \bar{K}_4\oplus \bar{K}_5 \oplus \bar{K}_6$.
Thus, the irreps that appear in a model with spinful $p_{x,y}$ orbitals on the honeycomb lattice are 
\begin{equation}
2\bar{\Gamma}_7\oplus \bar{\Gamma}_8\oplus\bar{\Gamma}_9 \text{ and }3\bar{K}_6\oplus\bar{K}_4\oplus\bar{K}_5
\label{eq:pxyirreps}
\end{equation}
As mentioned in the main text, they belong to a sum of three EBRs, $^1\bar{E}\uparrow G$, $^2\bar{E}\uparrow G$, and $\bar{E}_1\uparrow G$, induced from the $2b$ position. The double-valued EBRs for SG 183 are listed in Table~\ref{tab:183EBRs}.
(Since there is only one double-valued irrep of the little group at $M$, it cannot be used to distinguish EBRs and we do not need to consider it here.)
Since the $\bar{E}_1\uparrow G$ EBR is decomposable, generically, the band structure splits into four groups of bands.
One possibility is that one branch contains the irreps $\bar{\Gamma}_8$ and $\bar{K}_6$, another contains $\bar{\Gamma}_9$ and $\bar{K}_6$, another contains $\bar{\Gamma}_7$ and $\bar{K}_6$ and the last contains $\bar{\Gamma}_7$ and $\bar{K}_4\oplus \bar{K}_5$.
%If one branch contains $\bar{\Gamma}_8$ and $\bar{K}_6$ and another contains $\bar{\Gamma}_9$ and $\bar{K}_6$, then the third branch contains $\bar{\Gamma}_7$ and $\bar{K}_6$ and the fourth branch contains $\bar{\Gamma}_7$ and $\bar{K}_4\oplus \bar{K}_5$; 
In this case, each branch has the same irreps at high-symmetry points as an EBR listed in Table~\ref{tab:183EBRs}, but the EBR might come from orbitals on the $1a$ position.
In the language of Ref~\onlinecite{NaturePaper}, a branch that contains the same irreps as an EBR induced from orbitals located at a different site than the atoms is called an ``obstructed atomic limit.''
An obstructed atomic limit can have localized Wannier functions, but since those Wannier functions are not located where the atoms are located, a gap must close in order to reach the phase where the Wannier functions and the atomic orbitals are localized at the same sites.

The other possibility is that the bands disconnect in such a way that some branches do not have the same irreps as an EBR; this can happen if a branch contains $\bar{\Gamma}_8$ or $\bar{\Gamma}_9$ and $\bar{K}_{4}\oplus \bar{K}_5$.
A branch that does not have the same irreps as an EBR does not correspond to an atomic limit and cannot yield maximally localized Wannier functions that obey the crystal symmetry, centered at any position.

%According to Table~\ref{tab:183EBRs}, the set of irreps in Eq~\ref{eq:pxyirreps} is identical to those in a sum of four EBRs: 
%$\bar{E}_1\uparrow G$, $\bar{E}_2\uparrow G$, $\bar{E}_3\uparrow G$ induced from the $1a$ position and 
%$^1\bar{E}\uparrow G$ (or $^2\bar{E}\uparrow G$), induced from the $2b$ position. 

\begin{table}[h]
{\renewcommand{\arraystretch}{1.2}%
\centering
\begin{tabular}{c|c|c|c|c}
Wyckoff & EBR & $\bar{\Gamma}$ irreps & $\bar{K}$ irreps & No. bands\\
\hline
$1a$ & $\bar{E}_1\uparrow G$ & $\bar{\Gamma}_9$ & $\bar{K}_6$ & 2\\
$1a$ & $\bar{E}_2\uparrow G$ & $\bar{\Gamma}_8$ & $\bar{K}_6$ & 2\\
$1a$ & $\bar{E}_3\uparrow G$ & $\bar{\Gamma}_7$ & $\bar{K}_4\oplus \bar{K}_5$ & 2\\
$2b$ & $^1\bar{E}\uparrow G$ & $\bar{\Gamma}_7$ & $\bar{K}_6$ & 2\\
$2b$ & $^2\bar{E}\uparrow G$ & $\bar{\Gamma}_7$ & $\bar{K}_6$ & 2\\
$2b$ & $\bar{E}_1\uparrow G$ & $\bar{\Gamma}_8\oplus \bar{\Gamma}_9$ & $\bar{K}_4\oplus \bar{K}_5\oplus \bar{K}_6$ & 4
\end{tabular}}
\caption{Double-valued EBRs in SG 183, obtained from the BANDREP application.\cite{GroupTheoryPaper}}
\label{tab:183EBRs}
\end{table}

When time-reversal symmetry is imposed, we can compute the $\mathbb{Z}_2$ index.
Let us first consider the case when SOC is spin-conserving; following Ref~\onlinecite{NaturePaper}, we refer to this as ``Haldane'' SOC, as opposed to Rashba SOC, which, by our definition, is any SOC term that does not conserve spin.
An SOC term that conserves spin will also preserve inversion symmetry (the inversion operator is exactly the tensor product of the spinless $C_{2z}$ operator in Eq~(\ref{eq:unitarygensxy}) and the identity in spin space).
We can compute the $\mathbb{Z}_2$ index\cite{Fu2007} from the product of $C_{2z}$ eigenvalues at $\Gamma$ and $M$, since each band in the spinless model gives rise to a Kramers pair with SOC, whose inversion eigenvalue is the same as the spinless $C_{2z}$ eigenvalue.
% (this analysis does not require $|x|>0$.)
The eigenvalues of $C_{2z}$ are given in Tables~\ref{tab:Meigs} and \ref{tab:Geigs} and the product for each band is tabulated in Table~\ref{tab:C2zeigs} for all parameter regimes. (Note: there are three inequivalent $M$ points, but they share the same $C_{2z}$ eigenvalues).
%Whenever the band structure is gapped, which requires that either Eq~(\ref{eq:pxygap1}) or (\ref{eq:pxygap2}) is satisfied, the product of inversion eigenvalues is equal to $+1$.
%Hence, we deduce that in the presence of small SOC that preserves the bulk gap, the resulting system will have a trivial $\mathbb{Z}_2$ index.
The ordered lists of $C_{2z}$ eigenvalues in the last column of Table~\ref{tab:C2zeigs} reflect the particle-hole symmetry of our simplistic model; however, terms that break the particle-hole symmetry without inverting bands at $\Gamma$ or $M$ will not change the order of eigenvalues.
The fact that some particle-hole symmetric arrangements do not appear (namely, $+1,+1,+1,+1$ and $+1, -1, -1, +1$, is a surprising feature of our simple model.)
Since the product of $C_{2z}$ eigenvalues of the lowest band is always $-1$, whenever SOC opens a gap to the lowest energy band, it is a topological gap.
If there is a gap at half-filling, there are two possibilities: if the gap was open before SOC was added, then the parameters $t_{\sigma,\pi}$ are constrained by Eqs~(\ref{eq:pxygap1})--(\ref{eq:pxygap2}); comparing to Table~\ref{tab:C2zeigs} reveals that this gap will have a trivial $\mathbb{Z}_2$ index (consistent with the fact that a spinless model must have a trivial $\mathbb{Z}_2$ index.) On the other hand, in the parameter regime that violates Eqs~(\ref{eq:pxygap1})--(\ref{eq:pxygap2}), the spinless system will be gapless at half-filling; if SOC opens a gap, then Table~\ref{tab:C2zeigs} shows that the gap has a nontrivial $\mathbb{Z}_2$ index.

We now consider non-spin-conserving (Rashba) SOC and examine each of the high-symmetry points.
First, Rashba SOC cannot open a gap at $M$ since each spinless band is non-degenerate (when SOC is turned on, it will become a Kramers pair.)
Second, Rashba SOC cannot open a gap at $K$: if two bands are degenerate at $K$ in the spinless model, then they are in the $K_3$ representation, which we showed in the supplement of Ref~\onlinecite{NaturePaper} (Sec IIID) can only be gapped if the strength of Haldane SOC exceeds that of Rashba SOC.
Third, we examine the $\Gamma$ point. In the spinless model, bands come in degenerate pairs, which are described by the $\Gamma_{5}$ or $\Gamma_6$ representation, where $C_{2z}$ is represented by $\pm \mathbb{I}$.
When we consider spin, $C_{2z}$ is represented by $\pm \mathbb{I} \otimes i s_z$, where $s_z$ is the Pauli matrix describing the spin degrees of freedom.
A term that breaks spin conservation will not commute with $\mathbb{I} \otimes i s_z$.
Hence, no Rashba term can appear at the $\Gamma$ point because it will break $C_{2z}$ symmetry.
We conclude from examining all three high-symmetry points that any gap that is opened by small SOC is adiabatically related to a gap that is opened by spin-conserving SOC and hence the $\mathbb{Z}_2$ topological index obtained from inversion eigenvalues still holds.

\begin{table}[h]
\begin{tabular}{l|c}
Parameter regime & Prod. $C_{2z}$ eigs.\\
\hline
$\begin{aligned}
t_\pi& < -t_\sigma\\
t_\pi & <3t_\sigma\end{aligned} \Rightarrow \begin{cases} E_{\Gamma 2} < E_{\Gamma 1} \\ E_{M3}\!<\! E_{M1}\! <\! E_{M4}\! <\! E_{M2} \end{cases} $ & $-1,+1,+1,-1$\\[10pt]
\hline
$\begin{aligned}
t_\pi &> 3t_\sigma \\
t_\pi &< \frac{1}{3}t_\sigma \end{aligned} \Rightarrow \begin{cases} E_{\Gamma 2} < E_{\Gamma 1} \\ E_{M3,M4}<E_{M1,M2}  \end{cases} $ & $-1,-1,-1,-1$\\
\hline
$\begin{aligned}
t_\pi &< -t_\sigma \\
t_\pi &> \frac{1}{3}t_\sigma
\end{aligned} \Rightarrow \begin{cases} E_{\Gamma 2} < E_{\Gamma 1} \\ E_{M4}\! < \! E_{M2}\! < \! E_{M3}\! < \! E_{M1} \end{cases}$ & $-1,+1,+1,-1$\\
\hline
$\begin{aligned}
t_\pi &> -t_\sigma \\
t_\pi &> 3t_\sigma
\end{aligned} \Rightarrow \begin{cases} E_{\Gamma 1} < E_{\Gamma 2} \\ E_{M2}\! < \! E_{M4}\! < \! E_{M1}\! < \! E_{M3} \end{cases}$ & $-1,+1,+1,-1$\\[10pt]
\hline
$\begin{aligned}
t_\pi &> \frac{1}{3}t_\sigma \\
t_\pi &< 3t_\sigma \end{aligned} \Rightarrow \begin{cases} E_{\Gamma 1} < E_{\Gamma 2} \\  E_{M1,M2}<E_{M3,M4}  \end{cases} $ & $-1,-1,-1,-1$\\
\hline
$\begin{aligned}
t_\pi &> -t_\sigma \\
t_\pi &< \frac{1}{3}t_\sigma
\end{aligned} \Rightarrow \begin{cases} E_{\Gamma 1} < E_{\Gamma 2} \\ E_{M1}\! < \! E_{M3}\! < \! E_{M2}\! < \! E_{M4} \end{cases}$ & $-1,+1,+1,-1$
\end{tabular}
\caption{Product of $C_{2z}$ eigenvalues at $\Gamma$ and $M$ in order of increasing energy, for all possible parameter regimes.
According to Eqs~(\ref{eq:pxygap1}) and~(\ref{eq:pxygap2}), the spinless model can only be gapped in the parameter regimes corresponding to the second or fifth row.}
\label{tab:C2zeigs}
\end{table}

\section{$d_{z^2}$ and $d_{x^2-y^2}$ orbitals in $P4_232$}
\label{sec:208symmetries}

Since the lattice of $P4_232$ (SG 208) is primitive cubic, the lattice vectors are $\mathbf{e}_1=\hat{\mathbf{x}}, \mathbf{e}_2=\hat{\mathbf{y}}, \mathbf{e}_3=\hat{\mathbf{z}}$.
A unit cell is drawn in \MAINfigunitcell; we take the origin to be one of the corners of the cube.
Our model consists of $d_{z^2}$ and $d_{x^2-y^2}$ orbitals sitting at the corners and center of the cube.
These orbitals transform as representations of $T$ (generated by $C_{2z}$ and $C_{3,111}$), which is the ``site-symmetry group'' of the origin:\cite{ITA} that is, $T$ is the largest subset of $P4_232$ that leaves the origin invariant. 
The $d_{z^2}$ and $d_{x^2-y^2}$ orbitals transform as the one-dimensional irreps, $^1E$ and $^2E$, of $T$.\cite{Bilbao3}
Since they are time-reversed partners, they transform as a single irrep under the symmetries of $T$ and time-reversal.
(We also note that the tensor product of the spin-$\frac{1}{2}$ representation with $d_{z^2}$ and $d_{x^2-y^2}$ orbitals yields the two-dimensional irreps, $^1F_{3/2}$ and $^2F_{3/2}$.\cite{PointGroupTables})
Instead of choosing the diagonal set of matrix representatives, we choose a physically motivated basis, where a rotation by an angle $\theta$ about an axis $\hat{\mathbf{n}}$ is represented by $e^{i\theta \hat{\mathbf{n}}\cdot \mathbf{S}}$, projected onto the $d_{z^2}$ and $d_{x^2-y^2}$ orbitals. %, where $\mathbf{S}$ is the canonical set of $l=2$ spin operators.
This yields $C_{2z} = C_{2x} = C_{2y} = \sigma_0$, and
\begin{equation}
C_{3,111} = -\frac{1}{2}\mathbb{I} + \frac{\sqrt{3}}{2}(i\sigma_y) = \begin{pmatrix} -\frac{1}{2} & \frac{\sqrt{3}}{2} \\ - \frac{\sqrt{3}}{2} & -\frac{1}{2} \end{pmatrix}
\end{equation}

The screw operation $\tilde{C}_{2,110}\equiv \{ C_{2,110}|\frac{1}{2}\frac{1}{2}\frac{1}{2} \}$ mixes the two sublattices consisting of sites at the origin and sites at the center, $(\frac{1}{2},\frac{1}{2},\frac{1}{2} )$, of each unit cell.
The full representation of the symmetry operations is given by:
\begin{align}
U_{C_{2z}} &= U_{C_{2y}} = \tau_0 \otimes \sigma_0 \nonumber\\
U_{C_{3,111}}& = \tau_0\otimes (-\frac{1}{2}\sigma_0 + \frac{\sqrt{3}}{2}i\sigma_y )\nonumber\\
U_{\tilde{C}_{2,110},\mathbf{k}} &= e^{-\frac{i}{2}(k_x+k_y-k_z)}\tau_x\otimes \sigma_z,
\label{eq:208rep}
\end{align}
where the $\tau$ matrices act in the sublattice space.
Time-reversal is implemented by complex conjugation. Since the representation cannot be reduced without breaking time-reversal symmetry, it is ``physically irreducible.''\cite{Bilbao1}

\section{Symmetry constraints on the Wilson loop in $P4_2 32$ (SG 208)}
\label{sec:208wilsonsymm}

We show how the symmetries of $P4_232$ constrain the $z$-directed Wilson loop defined in \MAINeqwilsoncont ; we will always take the base point $k_{z0}=0$.
We will frequently utilize the transformation of the Wilson loop under a non-symmorphic unitary symmetry,
$\{ D_g |(\delta_x,\delta_y,\delta_z)\}$, such that $D_g$ acts in momentum space by $(k_\perp,k_z)\rightarrow (D_gk_\perp,-k_z)$, then
\begin{equation}
\mathcal{W}_{(D_gk_\perp,k_{z0})} = e^{2\pi i \delta_z}\tilde{U}_g(k_\perp,k_{z0}) \mathcal{W}^\dagger_{(k_\perp,-k_{z0})}(\tilde{U}_g(k_\perp,k_{z0}))^\dagger,
\label{eq:wilsnonsymm}
\end{equation}
where 
\begin{equation}
\left[\tilde{U}_g(\mathbf{k})\right]_{nm}\equiv  \langle u^n(D_g\mathbf{k}) | U_g |u^m(\mathbf{k})\rangle
\label{eq:utilde}
\end{equation}
This is a variation of Eq (B19) in Ref~\onlinecite{ArisCohomology} or Eq (D8) in Ref~\onlinecite{Wieder17}.
If $D_g$ does not invert $k_z$, then there is no dagger on $\mathcal{W}$ on the right-hand side of Eq~(\ref{eq:wilsnonsymm}).

%%%%%%%%%%%%%%
%%%%%%%%%%%%%%
%% LEFT OFF HERE REFERENCING SUPPLEMENT
%%%%%%%%%%%%%%
%%%%%%%%%%%%%%

\subsection{Wilson eigenvalues along $\bar{\Gamma}-\bar{X}-\bar{M}-\bar{Y}-\bar{\Gamma}$}
\label{sec:208wilsongapped}

Let us first consider the eigenvalues of the Wilson matrix at $\bar{\Gamma}$.
Enforcing $C_{2x}$ symmetry, (recall, $U_{C_{2x}}=\mathbb{I}_{4\times 4}$ from Eq~(\ref{eq:208rep})), Eq~(\ref{eq:utilde}) shows that $\tilde{U}_{C_{2x}}(\Gamma) = \mathbb{I}_{2\times 2}$; then Eq~(\ref{eq:wilsnonsymm}) yields $\mathcal{W}_{(\bar{\Gamma},0)} = \mathcal{W}^\dagger_{(\bar{\Gamma},0)}$.
This forces the eigenvalues of $\mathcal{W}_{(\bar{\Gamma},0)}$ to be real.
Enforcing $\tilde{C}_{2,110}$ symmetry (which has $\delta_z = \frac{1}{2}$), Eq~(\ref{eq:wilsnonsymm}) yields $\mathcal{W}_{(\bar{\Gamma},0)}=-\tilde{U}_{\tilde{C}_{2,110}}\mathcal{W}_{(\bar{\Gamma},0)}^\dagger \tilde{U}_{\tilde{C}_{2,110}}^\dagger$, which shows that the eigenvalues of $\mathcal{W}_{(\bar{\Gamma},0)}$ must be equal to $+1$ and $-1$.

Along the segment $\bar{\Gamma}-\bar{X}$, even without knowing $\tilde{U}_{C_{2x}}(\mathbf{k})$, we can utilize Eq~(\ref{eq:wilsnonsymm}) to deduce that the eigenvalues of $\mathcal{W}_{(k_x,0,0)}$ are equal to those of $\mathcal{W}_{(k_x,0,0)}^\dagger$. Consequently, the eigenvalues of $\mathcal{W}_{(k_x,0,0)}$ must either come in complex conjugate pairs or be real. However, since the eigenvalues of $\mathcal{W}_{(0,0,0)}$ are $+1$ and $-1$, the eigenvalues of $\mathcal{W}_{(k_x,0,0)}$ must also be fixed to $+1$ and $-1$ along the entire line, since there is no way for them to smoothly vary while satisfying the constraints of $C_{2x}$.

Applying the same logic along $\bar{X}-\bar{M}$ and $\bar{Y}-\bar{\Gamma}$ with $C_{2y}$ symmetry shows that the eigenvalues of $\mathcal{W}_{(\pi,k_y,0)}$ and $\mathcal{W}_{(0,k_y,0)}$ must also be pinned to $+1$ and $-1$.
Finally, applying $C_{2x}$ to $\bar{M}-\bar{Y}$ to the eigenvalues of $\mathcal{W}_{(k_x,\pi,0)}$, we deduce that the eigenvalues of the $z$-directed Wilson matrix are equal to $+1$ and $-1$ along the entire loop $\bar{\Gamma}-\bar{X}-\bar{M}-\bar{Y}-\bar{\Gamma}$.

\subsection{Protected band crossing along $\bar{\Gamma}-\bar{M}$}
\label{sec:208wilsoncrossing}

Applying Eq~(\ref{eq:wilsnonsymm}) with $\tilde{C}_{2,110}$ symmetry to the line $\bar{\Gamma}-\bar{M}$ shows that the eigenvalues of $\mathcal{W}_{(k,k,0)}$ are the same as the eigenvalues of $-\mathcal{W}_{(k,k,0)}^\dagger $; 
hence, the eigenvalues of $\mathcal{W}_{(k,k,0)}$ are either pure imaginary or come in pairs 
\begin{equation}
\lambda(k),-\lambda(k)^*
\label{eq:Wkk0eigs}
\end{equation}
We showed in Sec~\ref{sec:208wilsongapped} that the eigenvalues of $\mathcal{W}_{(0,0,0)}$ (and $\mathcal{W}_{(\pi,\pi,0)}$) are $+1$ and $-1$; this rules out the first possibility and hence the eigenvalues of $\mathcal{W}_{(k,k,0)}$ come in pairs $(\lambda(k), -\lambda(k)^*)$, which are degenerate when $\lambda(k) = \pm i$.
We now show that such a degeneracy can occur without any fine-tuning and that the parity of the number of degeneracies between $k=0$ and $k=\pi$ constitutes a topological invariant.

%, which rules out the first possibility and forces the Wilson matrix to take the form
%\begin{equation}
%\mathcal{W}_{(k,k,0)} = ie^{i\mathbf{a}(k)\cdot\sigma} = i\left( \cos a(k) + i\sin a(k) \left(\hat{n}(k)\cdot\sigma \right)\right)
%\end{equation}
%where $\mathbf{a}(k) = a(k)\hat{n}(k)$.
%The Wilson eigenvalues will be degenerate when $\sin a = 0$.
%We now show that such a crossing can occur without any fine-tuning and that the parity of the number of crossings constitutes a topological invariant.

To see this, we rely on an anti-unitary symmetry of the Hamiltonian:
$\mathcal{A} \equiv \mathcal{T}\tilde{C}_{2,110}^{-1}C_{2z}= \mathcal{T}\{C_{2,1\bar{1}0}|\bar{\frac{1}{2}}\bar{\frac{1}{2}}\frac{1}{2}\}$, satisfying $\mathcal{A}^2 = 1$.
Since $\mathcal{A}$ leaves $(k,k,k_z)$ invariant, the (antiunitary) generalization of Eq~(\ref{eq:wilsnonsymm}) is:
\begin{equation}
\mathcal{W}_{(k,k,k_{z0})} = -\tilde{U}_A(k,k,k_{z0})K \mathcal{W}_{(k,k,k_{z0})}K(\tilde{U}_A(k,k,k_{z0}))^\dagger,
\label{eq:Wdiag}
\end{equation}
where the minus sign comes from the fact that $\mathcal{A}$ includes a $\frac{1}{2}$ translation in the $\hat{z}$ direction ($e^{2\pi i \delta_z}=-1$), $K$ is the complex conjugation operator, and 
\begin{equation}
\left[\tilde{U}_A(\mathbf{k})\right]_{nm}\equiv  \langle u^n(\mathbf{k}) | \mathcal{A} |u^m(\mathbf{k})\rangle
\label{eq:atilde}
\end{equation}
Notice that $(\tilde{U}_AK)^2=\mathbb{I}_{2\times 2}$, from which it follows that $\tilde{U}_A(\mathbf{k}) = e^{ib_0(\mathbf{k}) + ib(\mathbf{k})(\cos\theta(\mathbf{k}) \sigma_x + \sin\theta(\mathbf{k}) \sigma_z)}$. Importantly, $\sigma_y$ does not appear in the exponential. Consequently $\tilde{U}_A$ is diagonalized by $\tilde{U}_A(\mathbf{k}) = e^{ib_0(\mathbf{k})}O(\mathbf{k})D(\mathbf{k})O(\mathbf{k})^T$, where $O(\mathbf{k})$ is a real orthogonal matrix and $D(\mathbf{k}) =  {\rm Diag}[e^{-ib(\mathbf{k})}, e^{ib(\mathbf{k})} ]$.
Defining 
\begin{equation}
W_k = O(k,k,0)^T\mathcal{W}_{(k,k,0)}O(k,k,0),
\end{equation}
Eq~(\ref{eq:Wdiag}) yields 
\begin{equation}
W_k = -D(k,k,0)W_k^* D(k,k,0)^*
\label{eq:wconstraint}
\end{equation}

Since $W_k$ has the same eigenvalues as $\mathcal{W}_{(k,k,0)}$, whose eigenvalues must come in pairs given by Eq~(\ref{eq:Wkk0eigs}), $W_k$ can be written as
\begin{equation}
W_k = ie^{i\mathbf{a}(k)\cdot\sigma} 
\end{equation}
for a smooth vector function $\mathbf{a}(k) = (a_x(k),a_y(k),a_z(k))$.
Eq~(\ref{eq:wconstraint}) then places the following constraints at each $\mathbf{k}$: 
\begin{align}
a_z\sin |\mathbf{a}| &= 0 = (a_y\sin b + a_x\cos b)\cos|\mathbf{a}|
\end{align}

Since $\mathbf{a}(k)$ is a smoothly varying function, there are two possibilities: either $\cos |\mathbf{a}(k)|=0$ for all $k$ (in which case the eigenvalues of $W_k$ are fixed to $\pm 1$) or $\sin |\mathbf{a}(k)|\neq 0$, $\cos |\mathbf{a}(k)|\neq 0$ except at isolated points, in which case, $a_x(k) \propto a_y(k), a_z(k)=0$.
%\begin{equation}
%W_k = ie^{ia(k)(\sin b(k)\sigma_x -\cos b(k)\sigma_y)},
%\label{eq:W1param}
%\end{equation}
%where $a(k)$ is a smooth, real-valued function of $k$.
(We rule out the case where $\sin |\mathbf{a}(k)|=0$ for all $k$ because it is inconsistent with the fact that the eigenvalues of $W_{k=0}$ are equal to $\pm 1$.)
The condition $a_x(k) \propto a_y(k), a_z(k)=0$ means that degeneracies in the spectrum of $W_k$ (and hence $\mathcal{W}_{(k,k,0)}$) occur when $a_x(k)=0$, which forces $a_y(k)=0$.
If such a degeneracy is present and linear in $k$, then it is not fine-tuned, in the sense that smoothly deforming $a_x(k)$ will move the degeneracy, but not remove it; such degeneracies can only be removed pairwise.

%In the case that Eq~(\ref{eq:W1param}) holds, eigenvalues of $W_k$ (and hence $\mathcal{W}_{(k,k,0)}$) are given by $ie^{\pm i a(k)}$, i.e., they are determined by a single smooth function of a single variable.
%It follows that linear degeneracies in the spectrum of $\arg W_k$ are not fine-tuned: if $a(k) = (k-k_0) + O((k-k_0)^2)$, then smoothly deforming $a(k)$ will move the degeneracy around, but will not remove it; in fact, such degeneracies can only be removed pairwise.

Since at both $\bar{\Gamma}$ and $\bar{M}$, the eigenvalues of $W_k$ are fixed to $+1$ and $-1$, the parity of the number of linear crossings is a topological invariant, that cannot be changed without closing the gap in the bulk band spectrum.
Returning to the possibility that $\cos |\mathbf{a}(k)|=0$ for all $k$: since in this case the eigenvalues of $W_k$ are never degenerate, it trivially follows that the parity of linear in $k$ band crossings cannot be changed without closing the bulk gap.

\subsection{Winding of the bent Wilson loop}
\label{sec:208bentwinding}

The product of $C_{2x}$ and $\tilde{C}_{2,110}$ yields the screw symmetry, $\tilde{C}_4 \equiv \{C_{4z} | \frac{1}{2}\frac{1}{2}\frac{1}{2} \}$.
Applying Eq~(\ref{eq:wilsnonsymm}) with $\delta_z = \frac{1}{2}$, and removing the dagger on the righthand side of Eq~(\ref{eq:wilsnonsymm}) because $\tilde{C}_4$ leaves $k_z$ invariant, requires that the eigenvalues of $\mathcal{W}_{(k,k,k_{z0})}$ are exactly opposite those of $\mathcal{W}_{(-k,k,k_{z0})}$.

If there is an odd number of linear band crossings in the spectrum of $\mathcal{W}_{(k,k,k_{z0})}$ for $0\leq k\leq \pi$, then one band must have eigenvalue $e^{i\varphi(k)}$, where $\varphi(0) = 0$ and $\varphi(\pi) = \pi$.
The eigenvalue of the other band is given by $e^{i\pi - i\varphi(k)}$, according to Eq~(\ref{eq:Wdiag}).
Then $\tilde{C}_4$ requires that one band of $\mathcal{W}_{(-k,k,k_{z0})}$ has eigenvalue $-e^{i\pi - i\varphi(k)} = e^{-i\varphi(k)}$; we use this band to define $\varphi(k)$ when $-\pi < k <0$, i.e., $\varphi(-k) = -\varphi(k)$. Thus, if we plot $\varphi(k)$ from $-\pi < 0 < \pi$, it ``winds'' from $-\pi $ to $\pi$.
This is exactly what is shown in \MAINfigwilsonz .
Applying the same logic to the other band shows that it winds in the opposite direction.

\section{Wilson-of-Wilson loop}
\label{sec:wilsofwils}

Let $C$ be the closed path in the surface Brillouin zone that traverses $\bar{\Gamma}-\bar{X}-\bar{M}-\bar{Y}-\bar{\Gamma}$ and, for each $k_\perp \in C$, let $|v_{1,2}(k_\perp)\rangle$ by the eigenstates of $\mathcal{W}_{(k_\perp,0)}$ with energies $-1$ and $+1$, respectively.
We define the Berry phase of the Wilson loop (the ``Wilson-of-Wilson'' loop) by $w = e^{i\oint_Cdk_\perp a(k_\perp)}$, where $a(k_\perp) =  i\langle v_1(k_\perp)| \partial_{k_\perp} |v_1(k_\perp)\rangle $.

We show that the symmetry $C_{2z}\mathcal{T}$ requires $w = \pm 1$.
In analogy to Eq~(\ref{eq:Wdiag}), the Wilson matrix satisfies,
\begin{equation}
\mathcal{W}_{(k_\perp,0)}=\tilde{U}_{C_{2z}\mathcal{T}}(k_\perp,0) K\mathcal{W}_{(k_\perp,0)}^\dagger K\tilde{U}_{C_{2z}\mathcal{T}}(k_\perp,0)^\dagger,
\end{equation}
where,
\begin{equation}
\left[ \tilde{U}_{C_{2z}\mathcal{T}} (k_\perp,k_z) \right]_{nm}\equiv \langle u^n(k_\perp,-k_z)| C_{2z}\mathcal{T} | u^m(k_\perp,k_z)\rangle
\end{equation}
When $k_\perp \in C$, $\mathcal{W}_{(k_\perp,0)}$ is Hermitian, as we showed in Sec~\ref{sec:208wilsongapped}.
Hence, $\tilde{U}_{C_{2z}\mathcal{T}}(k_\perp,0)K$ is an anti-unitary symmetry that commutes with $\mathcal{W}_{(k_\perp,0)}$ and hence does not mix the two Wilson bands, which are gapped with eigenvalues $\pm 1$ along $C$.
Thus, Eq~(\ref{eq:Wdiag}) can be applied with $\mathcal{W}$ replaced by $w$:
\begin{equation}
w=e^{i\phi(\bar{\Gamma})}KwKe^{-i\phi(\bar{\Gamma})}
\label{eq:wofw} 
\end{equation}
and
\begin{equation}
e^{i\phi(k_\perp)} = \langle v_1(k_\perp) | \tilde{U}_{C_{2z}\mathcal{T}}(k_\perp)  | v_1(k_\perp)\rangle,
\end{equation}
Eq~(\ref{eq:wofw}) shows that $w$ is real and equal to $\pm 1$.

\bibliography{NewTIs}